\documentclass[
	aps,
	noshowpacs,
	nofootinbib,
	eqsecnum,
	preprintnumbers,
	longbibliography,
	groupedaddress]{revtex4-2}
\usepackage{physics}
\usepackage{amssymb,amsmath,graphicx}
\usepackage{bm}
\usepackage{hyperref}
\usepackage{xcolor}
\usepackage{float}
\hypersetup{colorlinks=true,linkcolor=blue,citecolor=red,urlcolor=blue}
\usepackage[normalem]{ulem}

\definecolor{z1}{HTML}{11b9a5}
\definecolor{z2}{HTML}{d4a816}
\definecolor{z3}{HTML}{20a1c5}
\definecolor{z4}{HTML}{b1111e}

\newcommand{\mathperiod}{~.}

\begin{document}

\title{\bf Replicating Higgs fields in Ising gauge theory: the registry order.}

\author{Aleksandar Bukva}
\author{Koenraad Schalm}
\author{Jan Zaanen}
\affiliation{The Institute Lorentz for Theoretical Physics. $\Delta$-ITP \\
	Leiden University, Leiden, The Netherlands}

\begin{abstract}
We consider $Z_2$ gauge field theory coupled to ``Higgs'' matter fields invoking several copies of such matter, interacting entirely through the gauge fields, the 
$Z_2 \times Z_2 \times Z_2\cdots / Z_2$ and the $O(N) \times O(N) \times O(N) \cdots / Z_2$ families of theories. We discover that the Higgs phase of such theories is characterized by a hitherto unidentified ``registry'' order parameter. This is characterized by a gauge invariant $p = 2^{N_{\text{rep}}}/2$  Potts type symmetry  where $N_{\text{rep}}$ is the number of matter copies. The meaning of this registry is that the different matter copies align their vectors locally in strictly parallel or anti-parallel fashion, even dealing with the continuous $O(2)$ symmetry.  Supported by Monte-Carlo simulations, we identify the origin of this registry order in terms of the gauge interactions mediated by the fluxes (``visons'') associated with the $Z_2$ gauge fields, indirectly imposing the discrete symmetry in the gauge invariant global symmetry controlled effective order parameter theory. In addition, it appears that our simulations reveal a hitherto unidentified ``pseudo-universality'' associated with the very similar form of the overall phase diagrams of the various theories suggesting a remarkable ``governance'' by the gauge field part of the dynamics.  
\end{abstract}

\maketitle

\section{Introduction}
\label{intro}

The physics of gauge fields has its special traits and arguably the simplest incarnation of such an interacting theory, controlled by Ising ($Z_2$) local symmetry, has played a historical role in the subject, comparable to the (global) Ising model itself  in revealing generic principle. Early on, the pure $Z_2$ gauge theory was introduced by Wegner \cite{Wegner,Kogut}, revealing in a minimal setting the fundamentals in the form of the confining and deconfining phases. Much later it was gradually realized that, different from the confining phase, the deconfining phase is characterized by {\em topological} order \cite{Wen,Bais,Sondhi,visons,stripefrac,stripefracdemler}. Kitaev rediscovered this in the context of topological quantum computation in the form of his toric code \cite{Kitaev} demonstrating the non-Abelian braiding associated with its fluxes. This insight into the Ising deconfining state acquired in a recent era plays a central role in various guises in the modern topological order portfolio of condensed matter physics. 
  
In a next step one can couple in Higgs fields (``matter'') and generically a Higgs phase is formed besides the (de)confining phases.  Next to the role it played in quantum condensed matter (e.g. \cite{Sondhi,visons}), this flourished particularly in the context of liquid crystals. It was early on realized that the Higgs-confinement transition of $O(3)$ matter $Z_2$ gauge ($O(3)/Z_2$) is equivalent to the (uniaxial) nematic to isotropic transition \cite{toner95} (see also \cite{stripefrac,stripefracdemler}). Recently this was exploited to study ``generalized'' nematics characterized by the breaking of the non-abelian point groups in 3D \cite{nonabnematic}. 

In fact, this started with the pioneering work by Fradkin and Shenker \cite{Fradkinshenker} that focused in on the minimal case where matter is ``in the fundamental'', e.g. the Higgs field is governed by $Z_2$ symmetry as the gauge fields. They found that the ``maximally ordered'' Higgs phase and ``maximally disordered'' confining phase actually become {\em indistinguishable}. This confused the community for a long while and perhaps this got resolved for the first time by Huse and Leibler  \cite{HuseLeibler} invoking an analogy with amphiphilic phases as of relevance in e.g. biology. We will review this in more detail underneath since this motif will play an important role in the remainder. 
  
We stumbled on a natural extension of this discrete gauge theory portfolio. Although we have not managed to identify any circumstance where this could be of empirical relevance, as a theoretical construct it is entertaining. It adds some new general motifs, while shedding also an unexpected light on the basic physics of discrete gauge systems.   We call this ``discrete gauge theory with \textit{replicated} Higgs fields'', where various matter fields couple to the same gauge field. This defines a vast landscape of theories, and we will focus on the most elementary examples. We will limit ourselves to three (overall) dimensional systems characterized  by a single $Z_2$ gauge field coupling to ``replicated'' $Z_2$ and $O(2)$ Higgs fields.

Let us first define this ``replicated'' theory. Discrete gauge symmetry can only be handled departing from a UV lattice. Let us consider the Euclidean action that may be interpreted as a thermal problem in $D = d + 1$ dimensions, or either as the action of the quantum incarnation in $d$ space dimensions after Wick rotation. Consider a hyper-cubic lattice in $D$ overall dimensions and assign matter (Higgs) fields $\va{\phi}_i^a$  to the site $i$ with a symmetry that will be specified in a moment. The novelty is in the $a = [1,N_{\text{rep}}]$ different ``flavor'' copies of the matter fields. Next, assign gauge group matrices $U_{ij}$ to the \textit{links} between nearest-neighbor lattice sites. The action is then in full generality of the form,
\begin{widetext}
	\begin{equation}
		S[\{\phi^a,U_{ij}\}] = K \sum_{\square} \Tr [U_{12}U_{23}U_{34}U_{41}] + \sum_{a=1}^{N_{\text{rep}}} J_a \sum_{\langle ij \rangle} U_{ij} \va{\phi}_i^a \cdot \va{\phi}_j^a + \sum_{ab} J^L_{ab} \sum_i \va{\phi}_i^a \cdot \va{\phi}_i^b
	\end{equation}
\end{widetext}
in the usual guise of defining the gauge curvature in terms of the Wilson plaquette action (first term), coupling minimally to matter (second term). Once again, the only novelty is in the consideration of the $N_{\text{rep}}>1$ matter field copies. The last term corresponds with the minimal, gauge invariant couplings between the Higgs field ``replicas''. A crucial ingredient is that all replica’s are subjected to coupling to the same gauge field.

We will be focused on the minimal $Z_2$ gauge field incarnation in a thermal setting, addressing its statistical physics in 3 overall space dimensions. In terms of Ising spins $\pm 1$ with Pauli matrices  living on the bonds with operator $\tau_{ij}^z$,
\begin{widetext}
	\begin{equation}
		\label{eq:z2action}
		S_{Z_2}[\{\phi^a,U_{ij}\}] = -K\sum_{\square}\tau^z_{12}\tau^z_{23}\tau^z_{34}\tau^z_{41} - \sum_{a=1}^{N_{\text{rep}}}J_a\sum_{\langle ij \rangle}\tau^z_{ij} \va{\phi}_i^a \cdot \va{\phi}_j^a - \sum_{ab}J^L_{ab}\sum_i \va{\phi}_i^a \cdot \va{\phi}_i^b \mathperiod
	\end{equation}
\end{widetext}
For a single Higgs field this is just the thoroughly studied $O(\cdots)/Z_2$ action, where $O(\cdots)$ refers to the symmetry group of the matter field. We will concentrate on the elementary cases of the matter fields in the fundamental $O(\cdots) \rightarrow Z_2$ as well as the case of $O(2)$ matter.  

We repeat, different from the single copy versions, we have not managed to identify circumstances where these replicated theories become of relevance to experiment.  Perhaps the closest approach are the multi-component superconductors identified and analyzed by Babaev and coworkers \cite{Babaev1,Babaev2,Babaev3}. These would correspond with two $U(1)$ matter fields sharing the $U(1)$ gauging by electromagnetism. However, the latter is supposedly to be {\em non-compact} lacking magnetic monopoles: as will become clear, the analogous ``gauge fluxes'' (or ``visons'') of the $Z_2$ version are crucial to the physics we wish to discuss.

Not knowing quite what to expect, we explored this theoretical landscape in first instance through
Monte-Carlo simulations with a focus on establishing the nature of the phase diagrams. To establish the location and nature of the phase transition we employ standard methodology (thermodynamics, Binder criterium): details can be found in the appendix. 

This revealed surprises that we deem of sufficient interest to report here. These are already revealed by the simplest replicated $Z_2 \times Z_2 \times Z_2 \cdots / Z_2$ version. 
We first consider the limit where the local couplings between the matter fields (the $J^L_{ab}=0$ in Eq. (\ref{eq:z2action})) are absent for identical Higgs couplings $J_a=J~\forall~a$.  Although there are no direct couplings between the replicated matter fields, we identify a new, gauge invariant global symmetry. We call this the ``registry'' order parameter, associated with the {\em relative} orientation  of the local matter fields. As we will explain in Section \ref{replZ2}, for $N_\text{rep}$ replicated fields this is governed by a $p = 2^{N_{\text{rep}}} /2$ state Potts model. This symmetry  is automatically broken in the Higgs phase, while it restores in the confining phase. The consequence is that, different from the single copy version, this renders the Higgs and confining phase to be distinguishable as they are now
separated by a second order phase transition. The local couplings $J^L_{ab}$ break this symmetry explicitly turning these transitions into cross-overs, while  the onset of the ``registry'' phase transition in the gauge coupling ($K$) and matter coupling ($J_a=J$) can be manipulated by choosing matter couplings that are different for the copies $J_a\neq J_{b\neq a}$. 

We will then turn to the $O(2) \times O(2) \times \cdots/Z_2$ case (Section \ref{replO2}). For a single matter copy the Higgs phase is distinguishable from the confining phase since the former breaks symmetry spontaneously in the form of a (``spin'') nematic order parameter, involving for $O(2)$ a second order phase transition. However, invoking more than one copy we find that  this is characterized by the same registry order parameter as the $Z_2$ case for vanishing local couplings. The ramification turns out to be that the  confinement-to-Higgs (or isotropic-to-nematic) transition is lifted to a first order one for the reason that {\em two} global symmetries (registry and nematic) are simultaneously spontaneously broken at this transition.

\begin{figure*}[!h]
  \centering	\includegraphics[width=.42\textwidth]{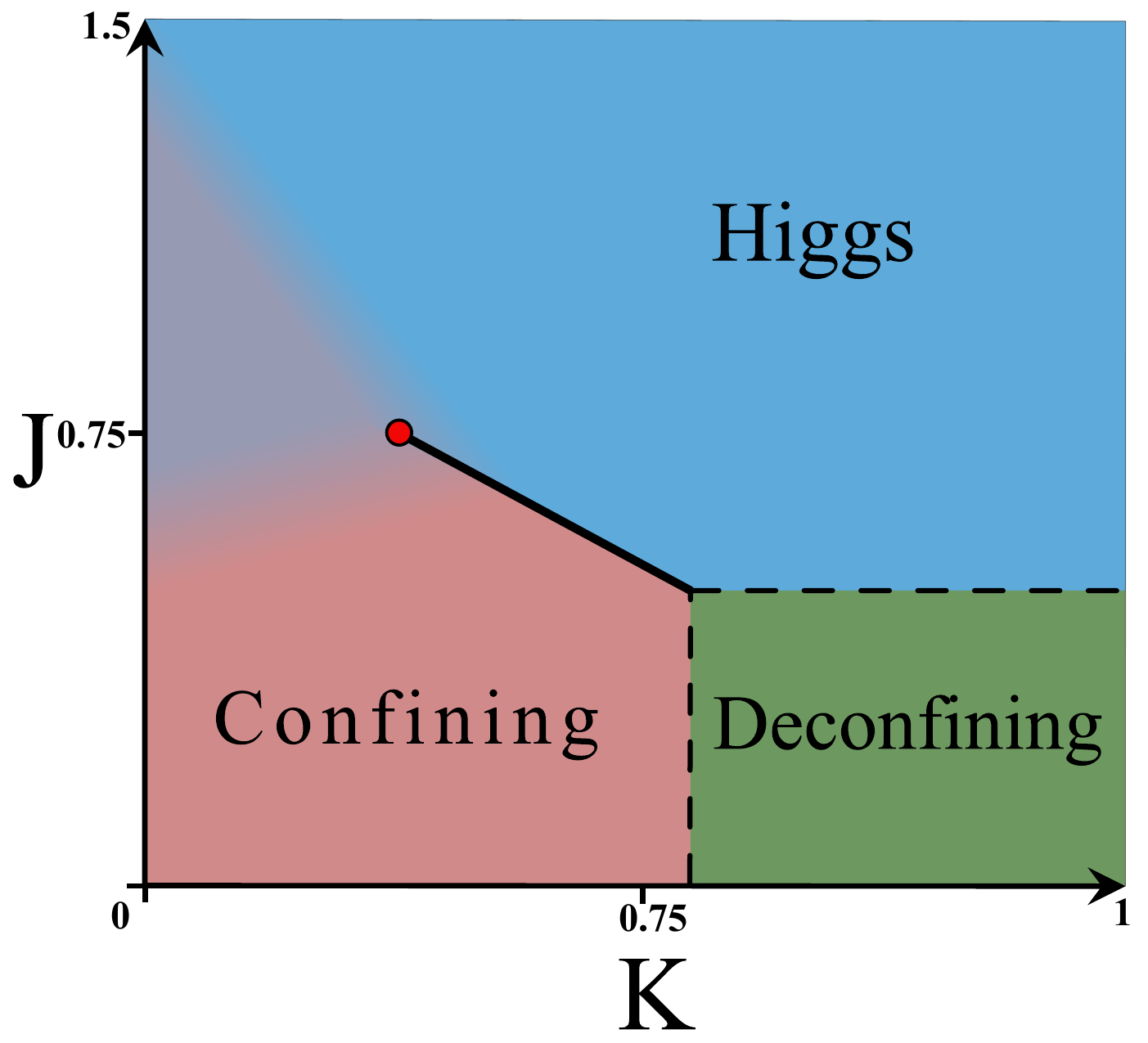}%
  \raisebox{-.2in}{
	\includegraphics[width=.45\textwidth]{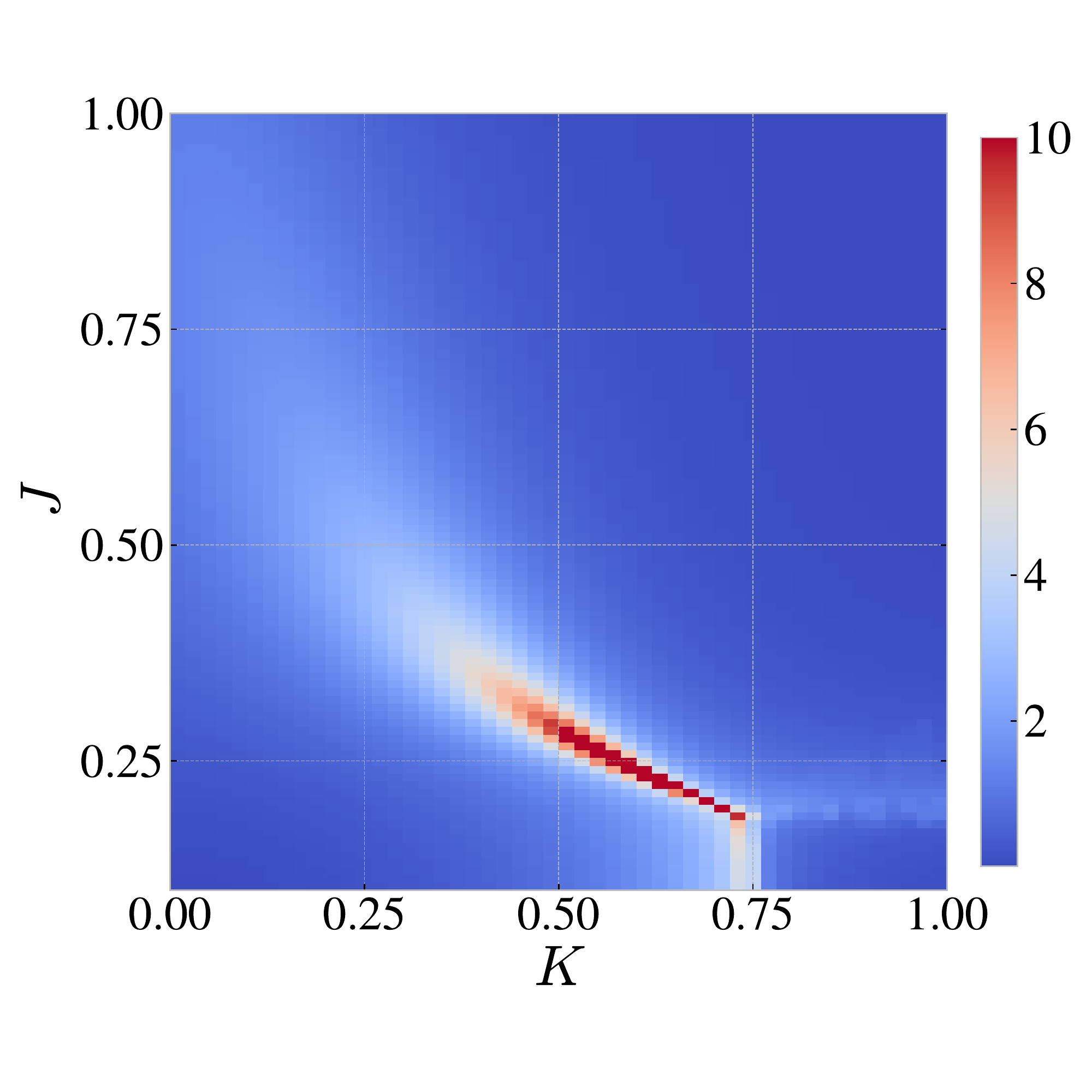}}
	\caption{The quantitative phase diagram as established by Monte-Carlo of the $Z_2 / Z_2$ theory as function of the gauge coupling $K$ and matter coupling $J$, see appendix \ref{app:mc-details} for details. The left panel gives the overview, further illustrated by  the outcomes for the specific heat in the right panel. The dashed- and drawn lines represent second- and first order transitions respectively, that we confirmed by a detailed analysis of the Binder cumulants. This is well understood (see main text): for small $J$ the matter sector is disordered and can be ignored, and one is just dealing with the pure gauge Wegner Ising gauge theory revealing the deconfining- and confining phases separated by a continuous phase transition, where the latter can be viewed as a condensate of gauge fluxes (``visons''). Upon increasing $J$ at  large $K$ one enters the ``Higgs-like'' phase that is famously indistinguishable from the confining phase -- a transition is absent as function of $J$ for small $K$. A peculiarity is the first order line emanating at the tricritical point anchored at the deconfinement transition. This reflects a van der Waals density driven liquid-vapor transition associated with the density of ``featureless'' domain walls, terminating at a critical end point.}
	\label{Z2Z2phasediagram}
\end{figure*}

Last but not least, it sheds further light on a peculiarity associated with the $Z_2$ gauge systems that we find to be more generic than anticipated. This departs from the structure of the phase diagram of the $Z_2/Z_2$ problem,  as function of gauge ($K$) and matter ($J$) coupling, Fig.\ref{Z2Z2phasediagram}. As pointed out already by Fradkin and Shenker \cite{Fradkinshenker}, by following the large $J$ (top side) and small $K$ (left side) evolution of the couplings no phase transition is encountered between the confining and Higgs ``phases'', stressing the indistinguishability. However, departing from the tricritical point associated with the deconfining phase one finds a strand of first order transition terminating at a critical endpoint. This was elucidated in full by Huse and Leibler \cite{HuseLeibler}, employing a dual language involving both the matter and gauge topological defects, that we will review first  to set the stage (Section \ref{Z2Z2review}). This is associated with a peculiar ``amplitude dynamics''  and we find that it is remarkably robust, being actually rather insensitive to the presence of gauge invariant global symmetries. This can already be seen in the elementary $O(2)/Z_2$ system where it appears to have been overlooked (see Fig. \ref{O2Z2phasediagram}) -- we are aware of only one publication where the first order ``strand'' was mentioned without further analysis \cite{Senthil}.  

We will end with a short discussion of our findings (section \ref{discussion}).

\section{ A short review of $Z_2$ gauge theory with matter.}
\label{Z2Z2review} 

Let us first remind the reader of the thorough understanding of Wegner Ising gauge theory, both for matter in the fundamental ($Z_2$) and the  case of matter fields with a larger symmetry than the fundamental. The pure gauge theory played a decisive role in the very early days of Yang-Mills theory by demonstrating the existence of confining/deconfining phase transitions \cite{Wegner} in 3 and higher (overall) dimensions, highlighted by the famous lecture notes  by Kogut \cite{Kogut}. Among others, the pure $Z_2$ gauge is dual to global Ising, while the deconfining phase was much later understood as being characterized by topological order. For a particularly appealing physical interpretation see the ``stripe fractionalization'' \cite{stripefrac,stripefracdemler}. 

The essence is the invariance of the theory Eq. \eqref{eq:z2action}  under the local gauge transformations at each site $i$, 
\begin{eqnarray}
	\ket{\text{state}} &\rightarrow  &\prod_j \sigma_{ij}^x \ket{\text{state}} \nonumber \\
	\phi_i^a &\rightarrow  & -\phi_i^a \mathperiod
	\label{gaugetrans}	
\end{eqnarray}
The bond variables are Ising valued $(\pm1)$ and the action is invariant under flipping the signs of all bonds emanating from any site $i$ when simultaneously the matter (vector) fields living on the site revers their signs. For vanishing (and by extension small) matter couplings $J_a$ one is in the realms of the pure gauge theory (see Fig.\ref{Z2Z2phasediagram}). This is best understood in terms of the topological excitations, the ``gauge fluxes'' \cite{Kogut} also called ``visons'' in the condensed matter literature \cite{visons}. The gauge invariant object is the $Z_2$ valued Wilson plaquette variable $\tau^z_{12}\tau^z_{23}\tau^z_{34}\tau^z_{41}$: for an even or uneven number of positive bond variables $\tau^z_{ij}$ this takes a positive or negative value, the latter representing gapped excitations when $K$ is large. 

\begin{figure*}[!h]
	\centering
	\includegraphics[scale=.5]{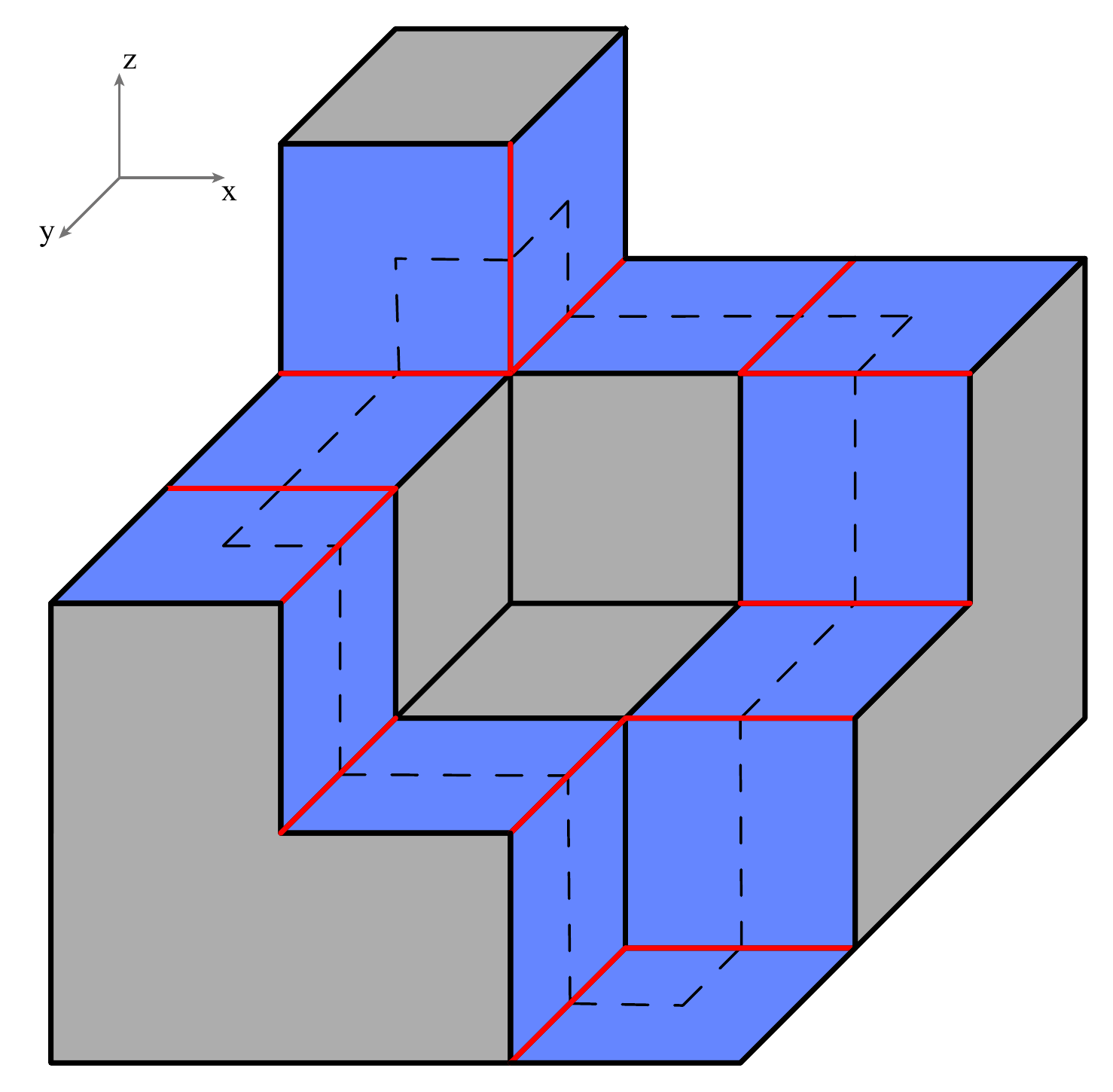}
	\caption{An example of a closed $Z_2$ gauge flux (``vison'') loop as these occur in the deconfining (and Higgs ``like'') phase representing the topological excitation of the gauge theory. The plaquettes indicated by blue represent a gauge invariant  flux $\tau^z_{12}\tau^z_{23}\tau^z_{34}\tau^z_{41} = -1$ immersed in a background of positive plaquette values. These have to form a connected line in three dimensions, the dashed line in this illustration. } 
	\label{loopexample}
\end{figure*}

These visons have a similar status as the monopoles of compact QED \cite{PolyakovQED}, the difference being that these fluxes have co-dimension $d-2$: in 3 dimensions these correspond with ``world lines'' (see Fig. \ref{loopexample}). It is easy to see that a Dirac seam emanates from the line (forming a surface) and for large $K$ these form small closed loops protecting the topological order. As for global $U(1)$ in 3D, upon reducing $K$ these loops grow in size to ``blow out'' at the transition to the confining phase that can be viewed as a condensate of the ``vison particles''. It is easy to demonstrate \cite{Kogut} that the (gauge invariant) Wilson loop exhibits a perimeter law when the visons are expelled from the vacuum (deconfinement), turning into an area law in the confining phase (vison condensate). This explains the small $J$ regime of all phase diagrams that we will present. 

\begin{figure*}[!h]
	\centering
	\includegraphics[scale=.4]{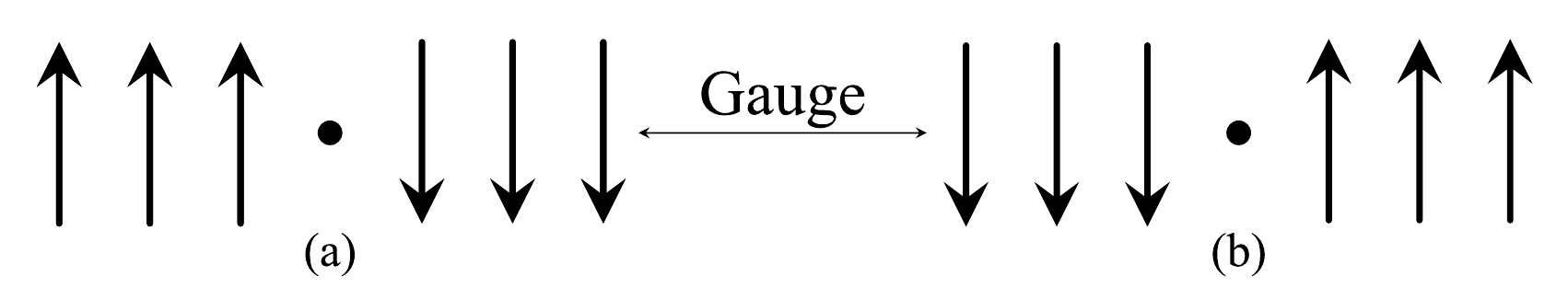}
	\caption{The featureless nature of the Higgs ``phase'' of the $Z_2/Z_2$ theory explained by inspecting the domain walls in the large $J,K$ regime. Depart from the unitary gauge (upper row) and construct a domain wall costing an energy $J$/unit cell. Restore the gauge invariance implying that on every site an up spin cannot be distinguished from a down spin. This shows that  the domain wall has the gauge invariant meaning of an object carrying energy but nothing else. This is the crucial insight behind the indistinguishability of the Higgs and confining phases.}
	\label{Z2Z2domainwall}
\end{figure*}

Let us now turn on the matter couplings $J^a$ to focus in on the case of minimal $Z_2$ matter ``in the fundamental''. As we already emphasized, the demonstration by Fradkin and Shenker \cite{Fradkinshenker} that the ``maximally orderly'' and ``maximally disorderly'' Higgs and confining phases are actually indistinguishable caused initially confusion. But in hindsight it is obvious, for the simple reason that a phase transition requires a {\em global}, gauge invariant symmetry to be broken. But in the presence of single $Z_2$ matter, global symmetry is erased. Consider the large $K,J$ limit; the visons are completely expelled and one can choose a unitary gauge fix taking all bond variables to be positive and the matter field living on the sites form an ordered Ising state with the spins, say, pointing up. However, according  the gauge transformation Eq. (\ref{gaugetrans}) on every site this can be swapped to a down spin. This sense of Ising order has no gauge invariant meaning and a global symmetry that is broken cannot be identified, and thereby there is no distinction from the confining phase. 

From tracking parameters along the large $J$ limit varying $K$ followed by descending along $J$ for small K one finds no phase transition proving the indistinguishability of the Higgs and confining phases. The oddity is however in the form of a ``strand'' of first order transitions emanating from the tricritical point anchored at the pure gauge confining-deconfining transition; the deconfining phase keeps of course its identity characterized by the expelled visons. This will be an important motif in the remainder: it is a peculiarity of the ``amplitude sector'' of  these gauge theories, that was elucidated by Huse and Leibler in a particularly appealing physical setting \cite{HuseLeibler}.

Although there is no manifest global symmetry, there is a ``material dynamics'' at work that becomes obvious considering the topological excitations, see Fig. \ref{Z2Z2domainwall}.  It is instructive to depart from the large $K,J$ limit. Consider the unitary gauge fix where one is dealing with a standard Ising model, characterized by domain walls as topological excitation costing an energy  $E \sim J$/unit cell. However, upon restoring gauge invariance it ``looses the symmetry'': for instance one can swap all spins to the left of the domain wall from up to down and the other way around to the right. Although the matter spins ``disappear'', the presence of the domain wall as en energetic excitation is a gauge invariant notion. This is the simple clue; upon reducing $J$ such featureless domain walls will start to proliferate.  

Let us now decrease $K$ such that (closed) vison loops start to form. It is easy to find out that the $Z_2/Z_2$ domain walls and visons relate to each other in the same way as the (Abrikosov) flux lines and magnetic monopoles of the compact $U(1)/U(1)$ theory, where the monopole ``cuts open'' the flux line. The vison loop ``cuts open'' the domain wall surface. For small $J$ and large $K$  in the ``Higgs like'' phase one finds accordingly large domain wall surfaces  with here and there a small hole (``vesicles'' \cite{HuseLeibler} ). However, for small $K$ and finite $J$ (``confinement like'') there are many visons  and accordingly the domain walls are ``cut in small pieces'' (``platelets''  \cite{HuseLeibler}).  

This offers the crucial insight regarding the origin of the first order line separating the confining and Higgs like regime. The net density of domain walls takes the role of density in the van der Waals theory dealing with liquid-vapor transitions. This is just the ubiquitous affair where the density changes discontinuous in a first order transition in the pressure-temperature diagram, terminating at a critical point. Huse and Leibler argue that this particular ``platelet'' versus ``vesicle'' incarnation is literally related to the behaviour of lipid membranes as of great relevance to e.g. biology\cite{HuseLeibler}. One take-home message of our work is that this peculiar ``amplitude dynamics'' is surprisingly ubiquitous in the whole landscape of theories defined by Eq. (\ref{eq:z2action}). 

The next general motif that will be important is associated with matter field characterized by a larger symmetry than the gauge sector. In general, this will imply that a gauge invariant {\em global} symmetry can be identified, that is broken in the Higgs phase, restoring the distinguishability with the confining phase. The simplest example is the single copy $O(N)/Z_2$ system. 

This ``left over'' global symmetry breaking is easy to infer in the large $K,J$ limit. In unitary gauge one finds here an ordered state breaking the $O(N)$ symmetry. However, the gauge transformation transforms the vector into minus itself. Consider $O(3)$ in 3D: this turns the vector in the {\em director} order parameter of a {\em uniaxial nematic}. The confinement-Higgs transition becomes in turn equivalent to the liquid crystal nematic-isotropic transition. This was used by Toner et al. \cite{toner95} to shed light on the origin of the well known first order character of this transition. One way is to consider small $K$ to integrate out perturbatively the gauge fluctuations, the outcome being that one recovers the Landau-de-Gennes theory governed by a rank two traceless symmetric tensor order parameter, allowing for a cubic invariant responsible for the first order transition. However, the dual (topological) language elucidates that {\em two} global symmetries are now simultaneously broken by the vison condensation as well as the manifest nematic order. We notice that this gauge theory strategy was used recently to completely classify and study ``generalized nematics''  associated with the breaking of rotational symmetry to any of the (non-Abelian) three dimensional point groups. This is a remarkably rich affair, with the highly symmetric point groups translating in high rank tensor order parameters \cite{nonabnematic,nonabnematic1,nonabnematic2}.

\begin{figure*}[!h]
  \centering	\includegraphics[width=0.425\textwidth]{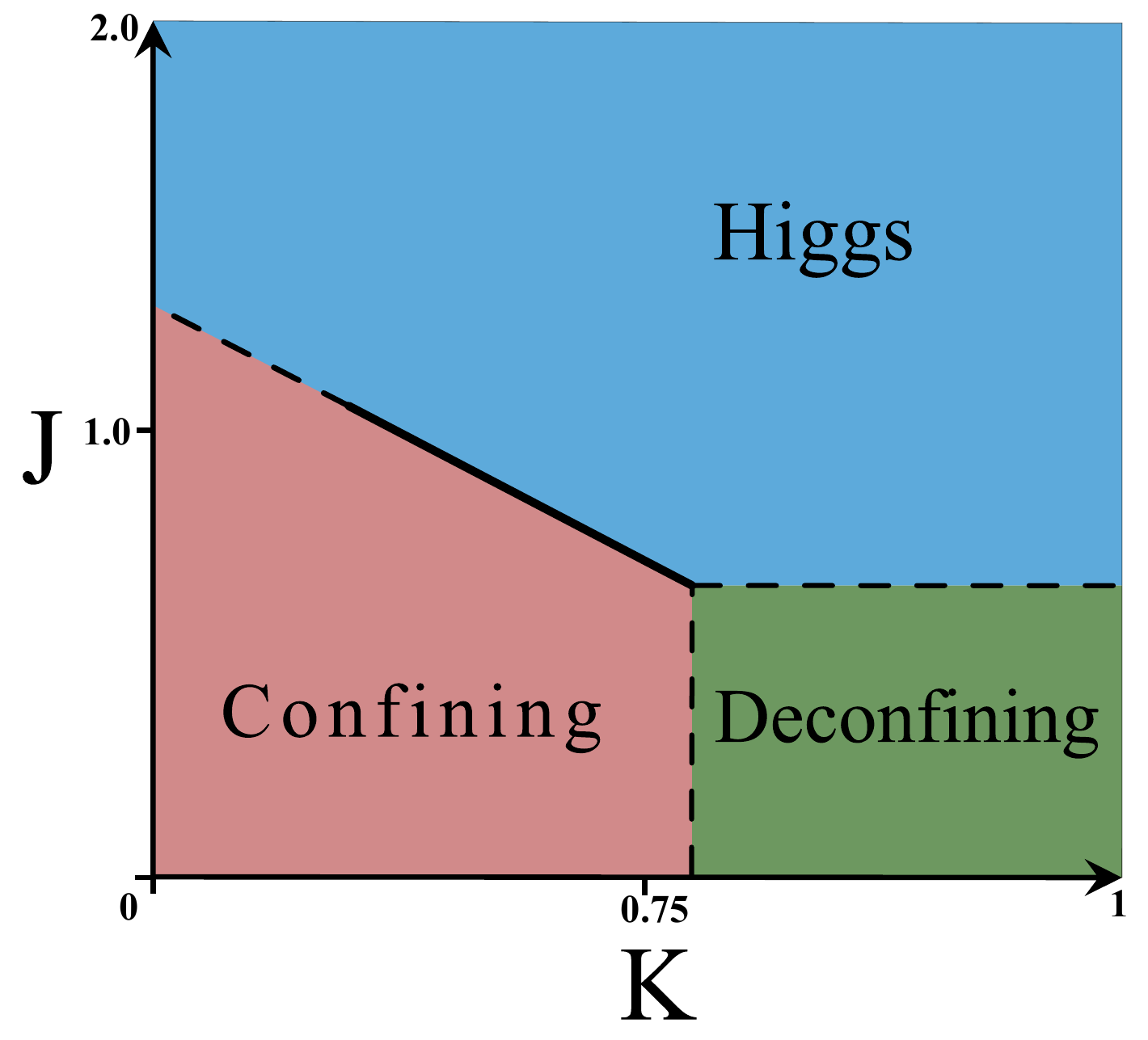}
  \raisebox{-.15in}{
	\includegraphics[width=0.44\textwidth]{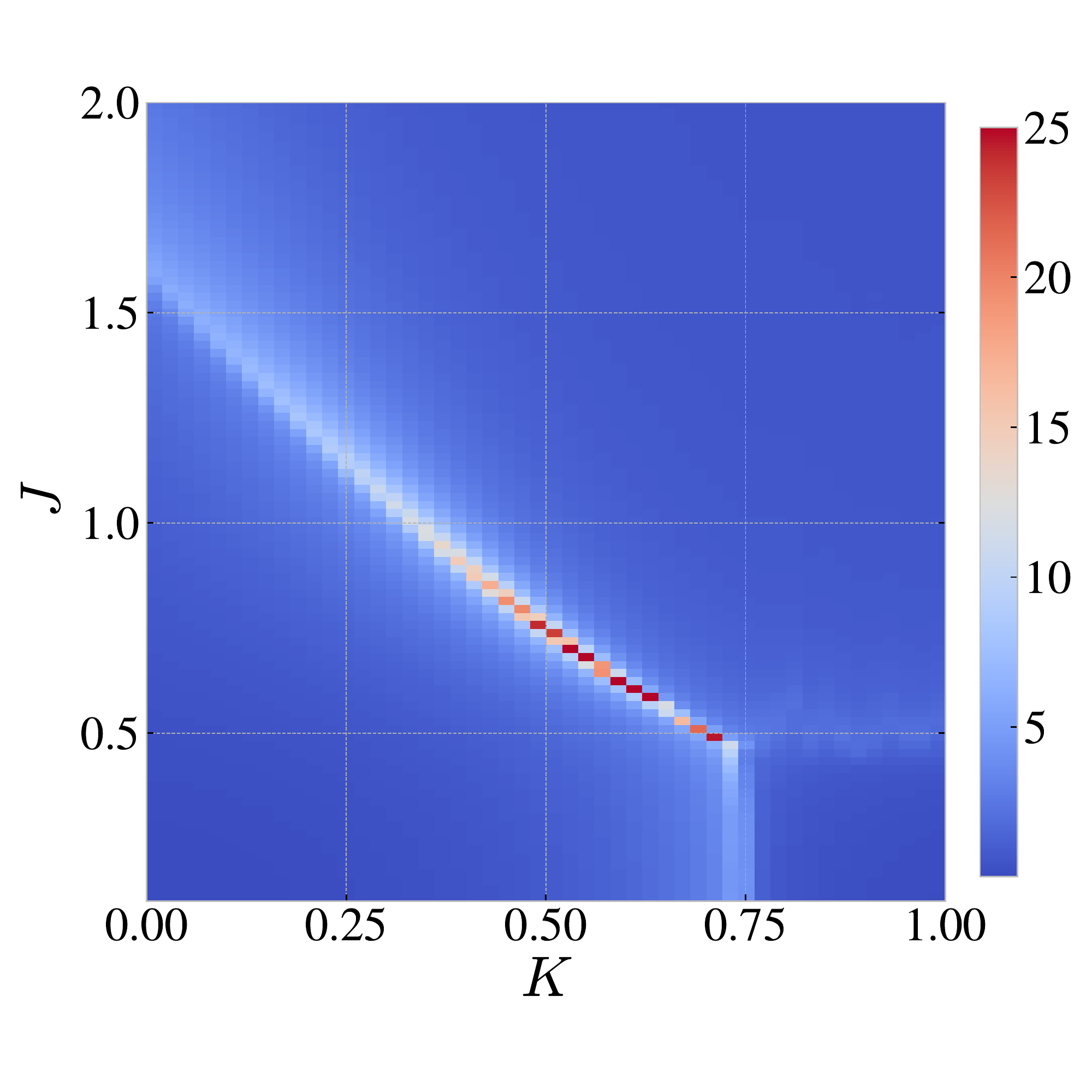}}
	\caption{In the left panel the phase diagram of the $O(2)/Z_2$ theory, established from the Monte Carlo results for the specific heat $c_V$ in the right panel. This can be directly compared with the $Z_2/Z_2$ theory, Fig.\ref{Z2Z2phasediagram}. One infers that it looks very similar, the only difference being that now the ``gap'' between the termination of the first order line and the $K \rightarrow 0$ limit between the confining- and Higgs phase is now interrupted by a second order phase transition associated with the breaking of the gauge invariant ``half-periodic'' $O(2)$ symmetry. The fact that in other regards the phase diagrams look so similar is surprising, see the main text. } 
	\label{O2Z2phasediagram}
\end{figure*}

A final motif that will be useful in the remainder is associated with the ``outlier'' Abelian matter $O(2)/Z_2$ case that is also well known, especially in the context of ``fractionalized'' superconductivity (e.g., \cite{visons}).  Different from the non-Abelian $O(N)$ cases in the small K limit the rank two tensor order parameter simplifies to a simple $O(2)$ with halved periodicity, 
 \begin{equation}
 H_{\mathrm{eff}, K \rightarrow 0} = -J' \sum_{\langle i, j \rangle} \left( \vec{\Phi}_i \cdot \vec{\Phi}_j \right)^2  = -J' |\Phi|^2 \sum_{\langle i, j \rangle} \cos \left( 2 (\phi_j - \phi_i ) \right) ~,
 \label{directorH}
 \end{equation}         
using $\vec{\Phi}_i = |\Phi | e^{i \phi_i}$. This has the obvious ramification that for $K \rightarrow 0$ this transition continues to be second order.  In Fig.\ref{O2Z2phasediagram} we show the phase diagram. There is now indeed a second order transition separating the Higgs and confining phases, associated with the disappearance of the ``halved periodicity'' XY order parameter characterizing the Higgs phase. However, upon increasing $K$ this turns into a first order line again. Strikingly, this first order ``strand'' is even quantitatively very similar as its analogue in the $Z_2/Z_2$ case: the main difference is just that the ``connection'' between confining and Higgs is now closed off by the transition involving the manifest XY order parameter.  This is clearly associated with the amplitude of the order parameter that obviously submits to the same ``Huse-Leibler'' logic, being controlled by the density of the ``non-topological'' matter defect ``fragments'' near the tricritical point. 

Although it has been previously observed that close to the tricritical point the Higgs-confinement transition has turned first order \cite{Senthil} it took us by surprise that the $O(2)/Z_2$ behavior behaves so similarly as to the $Z_2/Z_2$ case: the only essential difference is that the ``gap'' between the critical end point and the $K \rightarrow 0$ limit is just ``filled'' with the nematic-to-isotropic like second order transition, barely affecting even the locus of the end point of the first order line. A priori it is not at all obvious why this is so similar. For instance, the matter topological excitations, as identified in unitary gauge, are now XY-vortices with a quantized rotation associated with the halved periodicity. These are lines (and not surfaces) in 3D, in stark contrast with the Ising domain walls of the $Z_2/Z_2$ theory. We will encounter underneath other variations on this theme, invariably revealing this somewhat mysterious quantitative universality of the Huse-Leibler motif.     

\section{Replicating the $Z_2$ matter fields: two copies.}
\label{replZ2}

After these preliminaries let us now turn to the main subject: the family of ``replicated'' Ising gauge theories. The simplest case is the $Z_2$ matter theory with two matter copies:  $Z_2 \times Z_2 / Z_2$. In fact, the most interesting  case is the truly minimal one where we take the same matter field coupling $J_1 = J_2 = J$ and set the local couplings to zero $J^{12}_L = 0$. Under these circumstances the matter fields couple only through the Ising gauge fields.  

\begin{figure*}[!h]
  \centering	\includegraphics[width=0.425\linewidth]{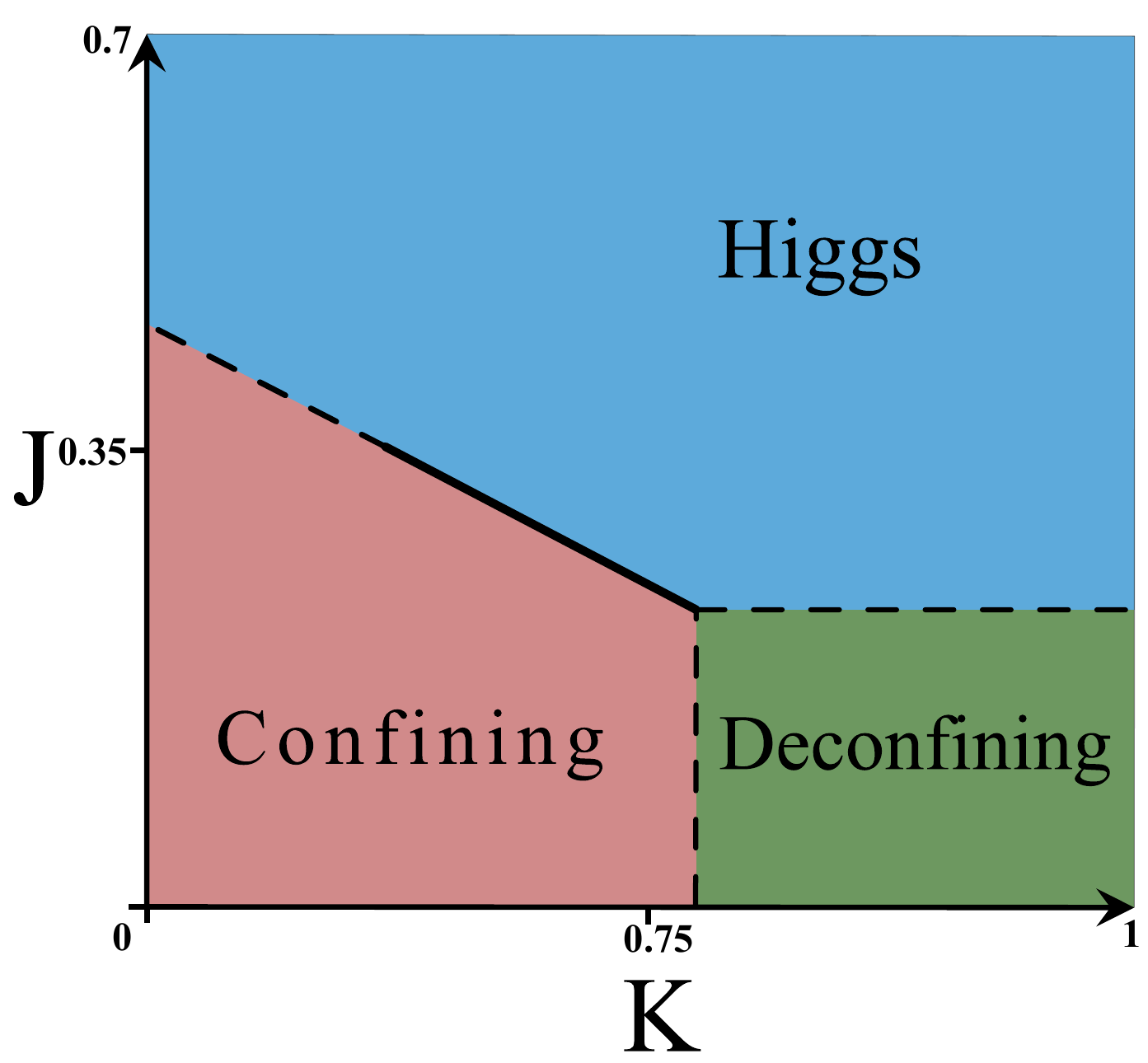}
  \raisebox{-.15in}{
	\includegraphics[width=0.44\linewidth]{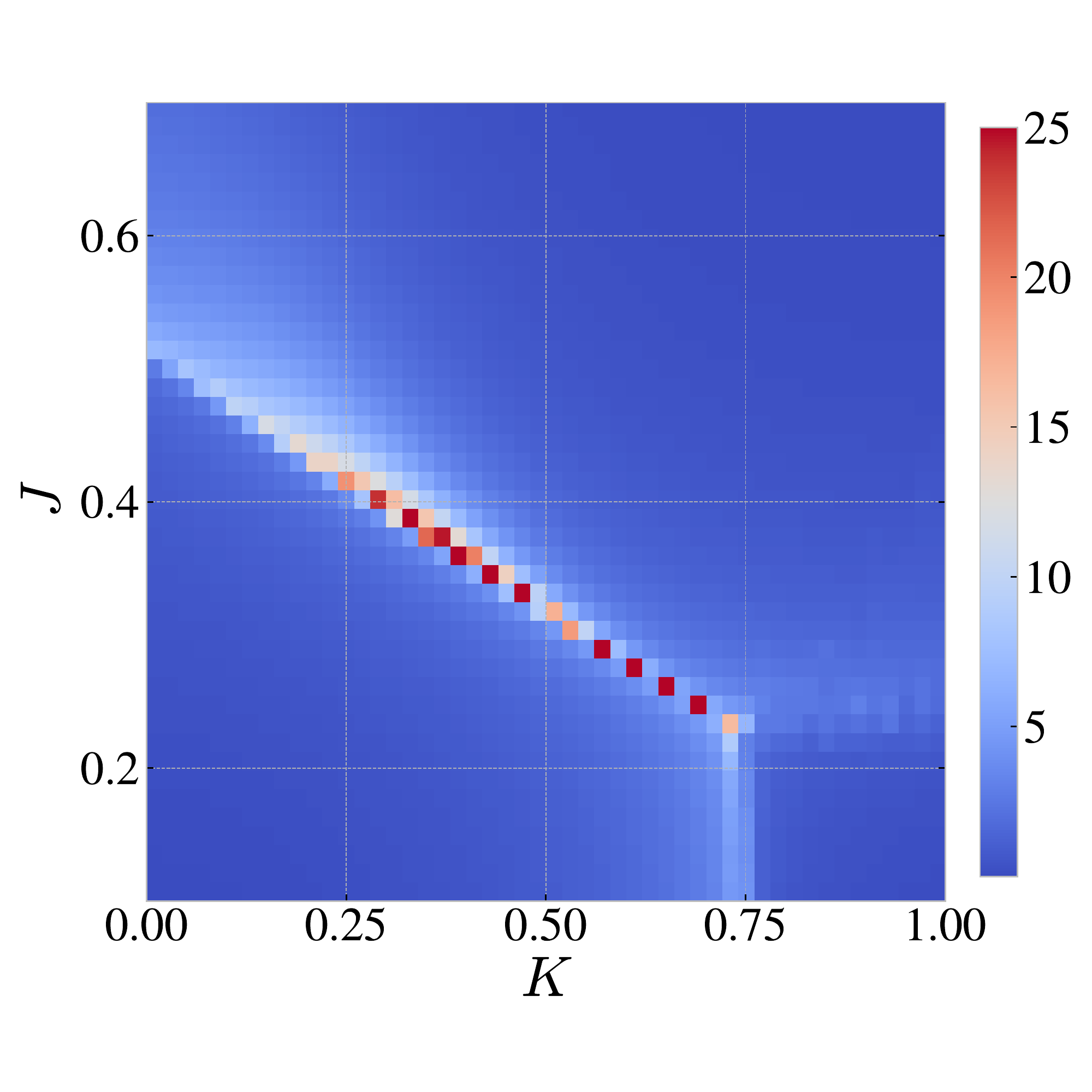}}
	\caption{ The phase diagram of the ``two copy''  $Z_2 \cross Z_2 / Z_2$ theory for $J_1 = J_2  = J$ and $J_L^{12} = 0 $ illustrated by  the Monte Carlo results  for the specific heat $c_V$ in the right panel. This looks very similar as to the $O(2)/Z_2$ case (Fig.\ref{O2Z2phasediagram}) in turn similar to the $Z_2/Z_2$ case  except for the connection between Higgs- and confinement is closed off by an honest second order transition. As explained in the text, one can now identify an Ising valued gauge invariant ``registry'' order parameter that breaks the symmetry spontaneously in the Higgs phase having a similar role as the nematic order parameter of the $O(2)/Z_2$ case.} 
	\label{Z2xZ2Z2phasediagram}
\end{figure*}

Let us start with the simplest example of a $Z_2$ gauge theory with replicated matter: the $Z_2 \times Z_2 / Z_2$ case. In Fig.\ref{Z2xZ2Z2phasediagram} we show the phase diagram as established by our Monte-Carlo simulations.  We infer that this is a very close sibling of the $O(2)/Z_2$ phase diagram that we just discussed Fig.\ref{O2Z2phasediagram}. Next to the ubiquitous Higgs-deconfining transition, the transitions between the Higgs- and confining phase looks very similar, including the first order line emanating from the tricritical point turning second order roughly at the locus of the $Z_2/Z_2$ critical endpoint. 

What is going on here? As explained earlier, for the transition between confining and Higgs at low $K$ to be a true second order phase transition, there has to be a broken global symmetry. In fact this is indeed controlled by a gauge invariant order parameter with a global Ising symmetry that may not be directly obvious to the reader, but it is quite simple. Consider again the unitary gauge with all bond spins $+ 1$. Because we have two matter fields, we are now dealing with two {\em independent} Ising spin systems living on the sites that will both be ordered for large $J,K$. As we emphasized in Section \ref{Z2Z2review}, after restoring the gauge invariance  both Ising spin systems ``loose their symmetry'' according to the $Z_2/Z_2$ rule book. But now we observe that the relative orientation of these two spin systems actually corresponds  with a  gauge invariant, global $Z_2$ symmetry! 

\begin{figure*}[!h]
	\centering
	\includegraphics[scale=.4]{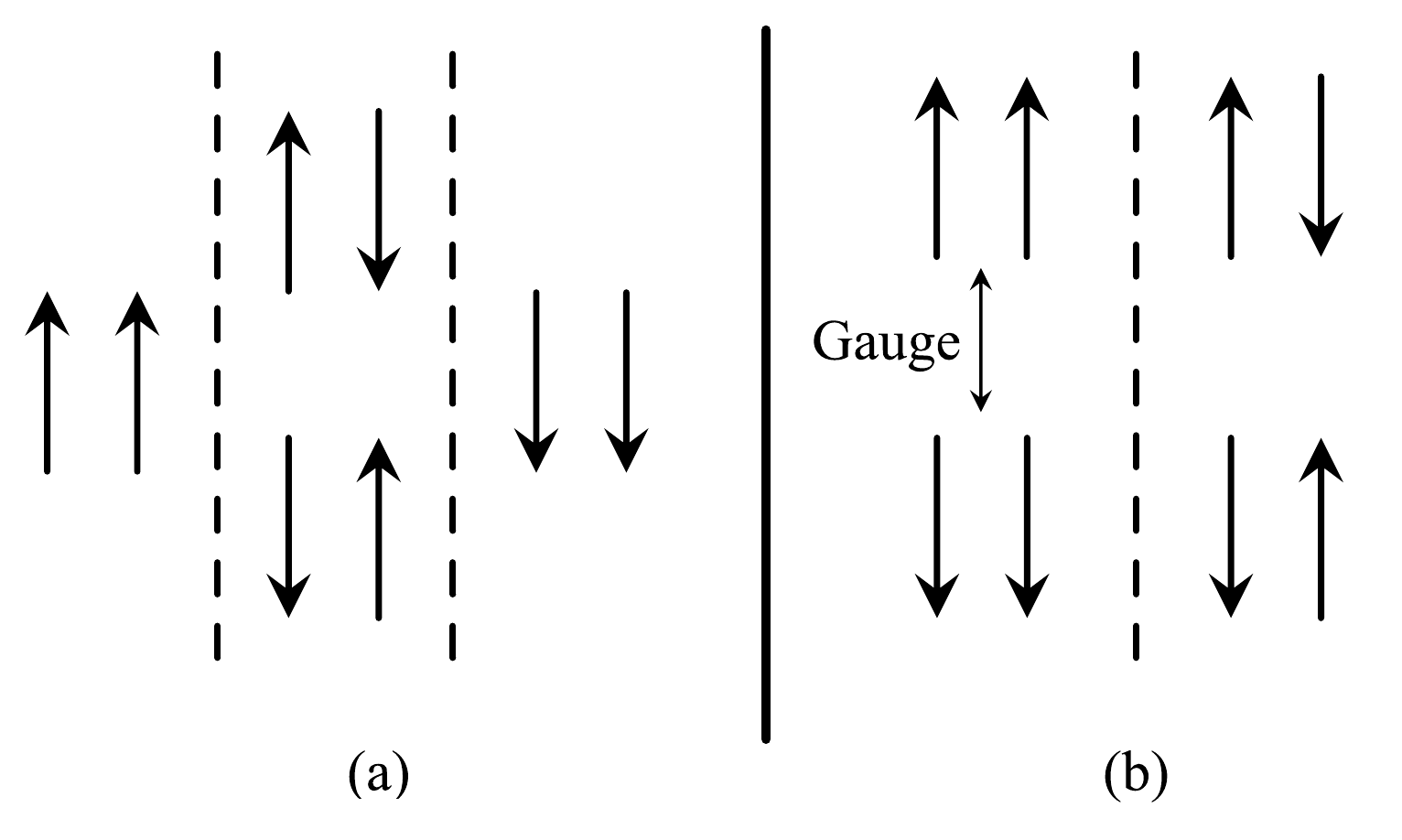}
	\caption{The ``registry'' order parameter of the two copy $Z_2 \times Z_2/Z_2$ theory. Depart from the large $J,K$ limit and take a unitary gauge fix as in Fig.\ref{Z2Z2domainwall}. One can now construct two types of domain walls associated with the two copies (upper line). However, upon restoring gauge invariance only two of the four on-site configurations can be distinguished: the spins are locally either parallel- or anti-parallel.  This is the global $Z_2$ symmetry that is broken in the Higgs phase.} 
	\label{twocopyregistry}
\end{figure*} 

Take the matter spins ``1'' to be pointing in the positive direction, and the ``2'' spins can be either parallel or anti-parallel to the ``1'' spins. Let us now see what happens with this ``registry'' upon restoring the gauge invariance, see Fig.\ref{twocopyregistry}. Consider parallel registry on a particular site like  $ \uparrow_{1} \uparrow_{2}$ and  under a gauge transformation both spins flip -- a single sector $\uparrow, \downarrow$ is not  gauge invariant. But it follows that the {\em relative} orientation of the matter spins in both sectors can be either {\em parallel} or {\em anti-parallel} and after restoring the gauge invariance it continues to be distinguishable whether one is dealing with locally parallel or anti-parallel configurations:  $ \downarrow_{1} \downarrow_{2} \leftrightarrow \uparrow_{1} \uparrow_{2}$ versus $ \uparrow_{1} \downarrow_{2} \leftrightarrow \downarrow_{1} \uparrow_{2}$! This is what we call the ``registry order parameter'' which carries clearly a global $Z_2$ charge!

\begin{figure*}[!h]
	\centering
	\includegraphics[scale=.7]{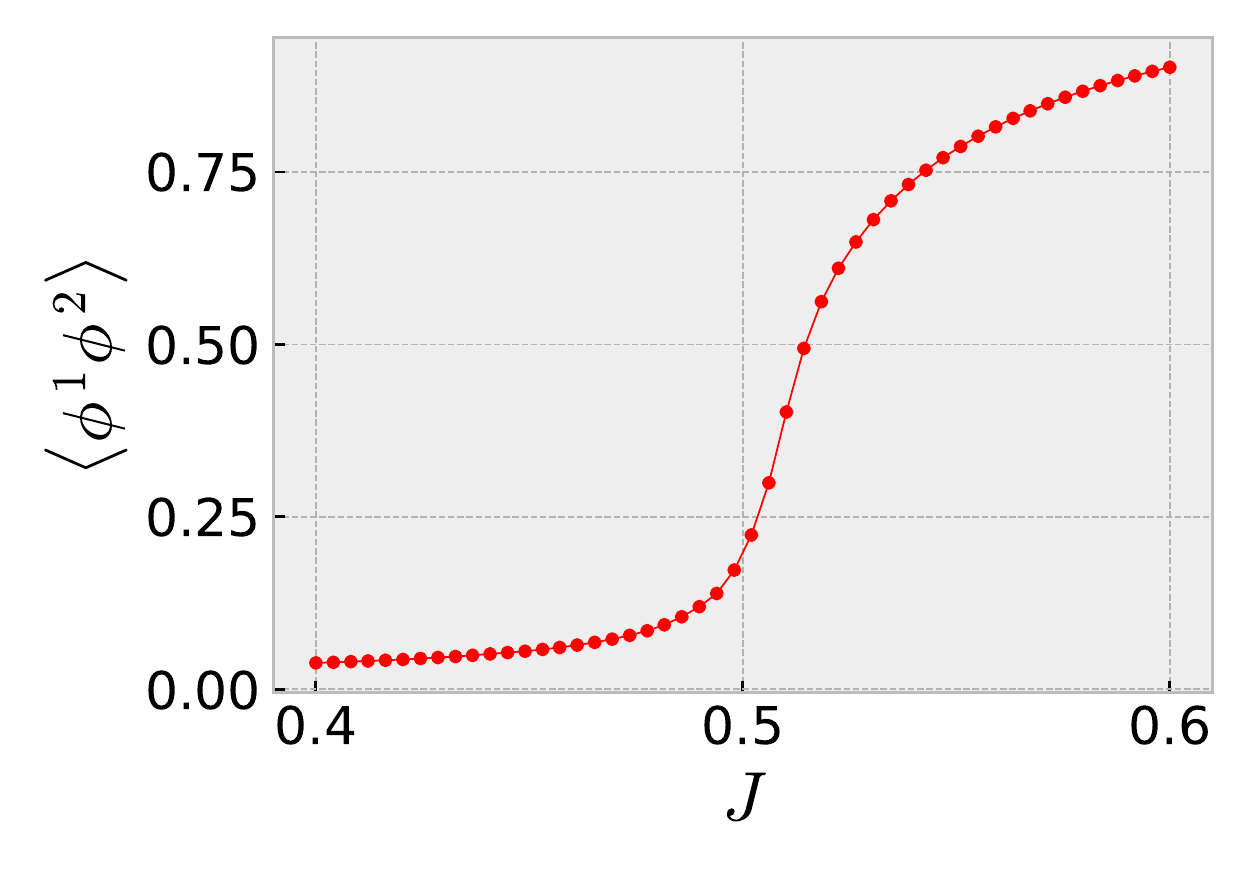}
	\caption{Expectation value of the registry order parameter $\langle \phi^1\phi^2\rangle$ of the  $Z_2 \times Z_2 / Z_2$ theory, along the $K=0$ slice.} 
	\label{Z2xZ2Z2orderParameter}
\end{figure*}

This ``registry'' symmetry breaks spontaneously in the Higgs phase, causing a two fold degenerate ground state: either the ``parallel'' or ``anti-parallel'' registry takes over. This registry order parameter $\langle \phi^1\phi^2\rangle$ is easy to measure and it is precisely what we find in the Monte-Carlo simulation of the Higgs phase, see Fig.\ref{Z2xZ2Z2orderParameter}.

At first sight this may be a bit confusing given that the matter fields interact only via the gauge couplings. However, the simple logic in the previous paragraph just reveals that this replicated system has in the Higgs phase a doubly degenerate ground state associated with ``parallel'' and ``anti-parallel'' registry. It is easy to construct gauge invariant domain walls between domains with opposite registry, as the ``featureless'' domain walls of the $Z_2/Z_2$ theory now acquire the gauge invariant meaning that they represent a jump in the registry order. Notice that the phase diagram is in all regards other than the second order  Higgs-confinement transition a near quantitative copy of the $Z_2/Z_2$ phase diagram. Given the lessons of the $O(2)/Z_2$ theory this is perhaps not surprising since the ``microscopy'' of the $Z_2 \times Z_2 / Z_2$ theory is a close cousin of the  $O(2)/Z_2$ case.  

\begin{figure*}[!h]
	\centering
		\includegraphics[width=.32\textwidth]{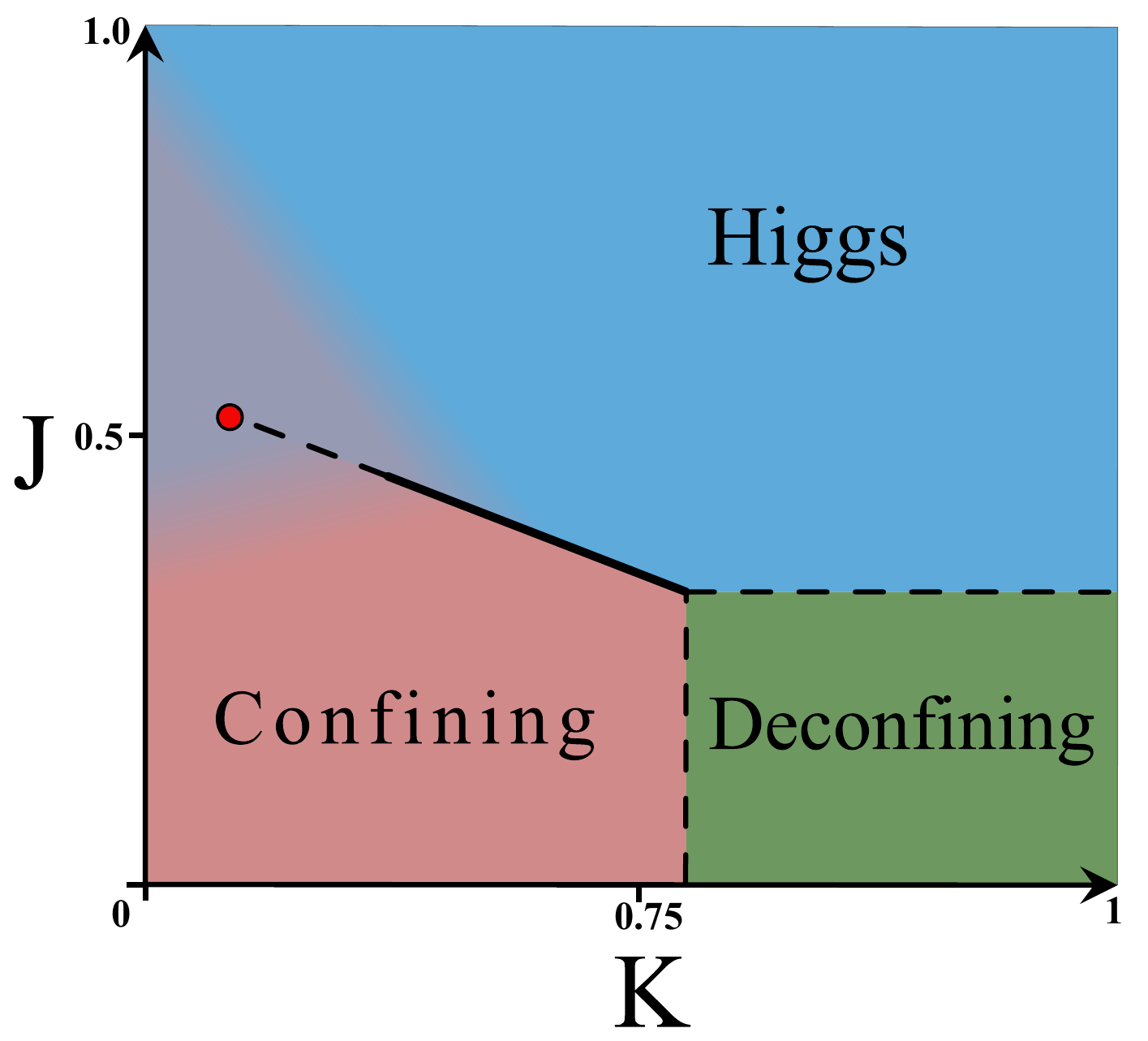}
		\includegraphics[width=.32\textwidth]{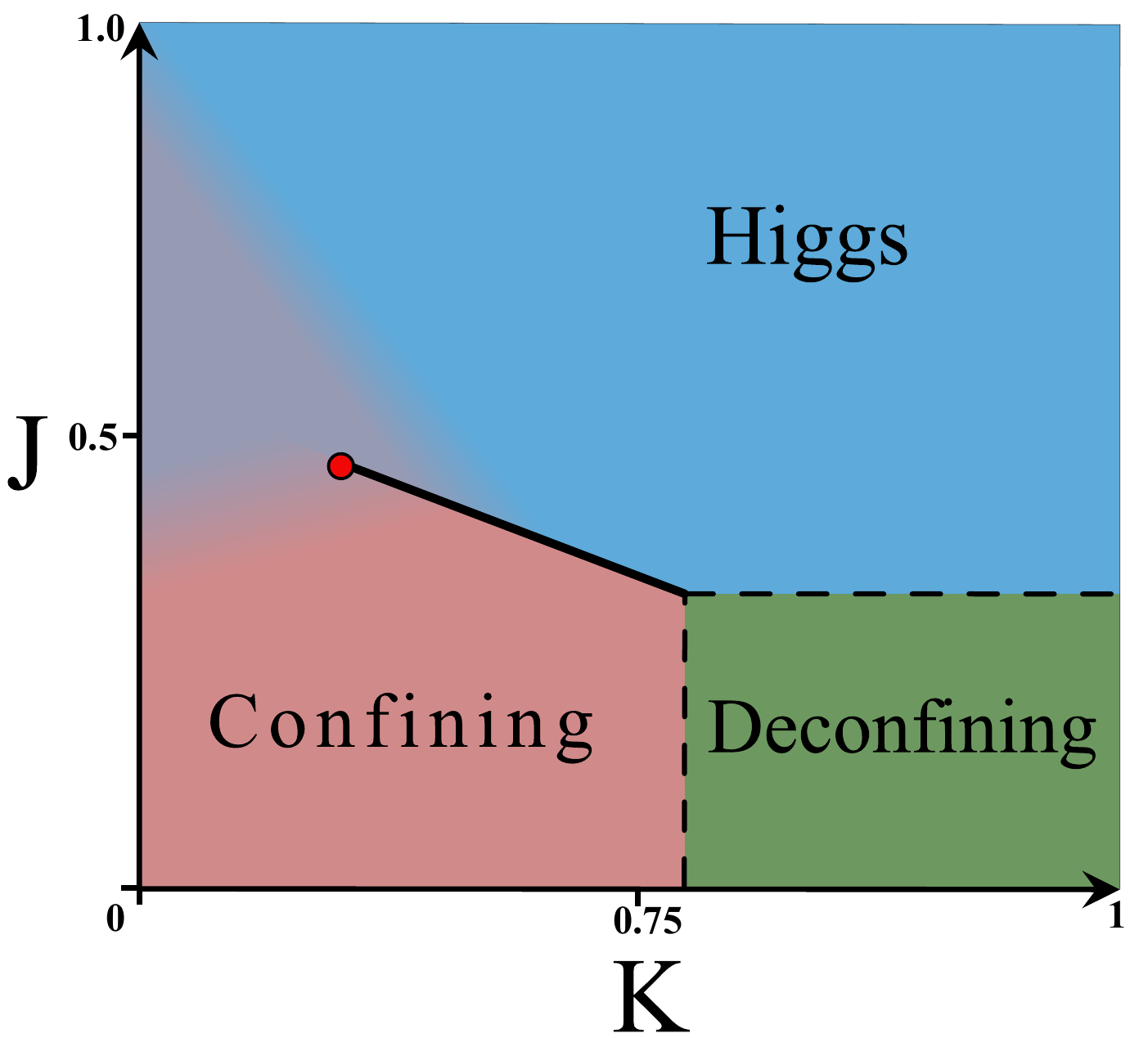}
		\includegraphics[width=.32\textwidth]{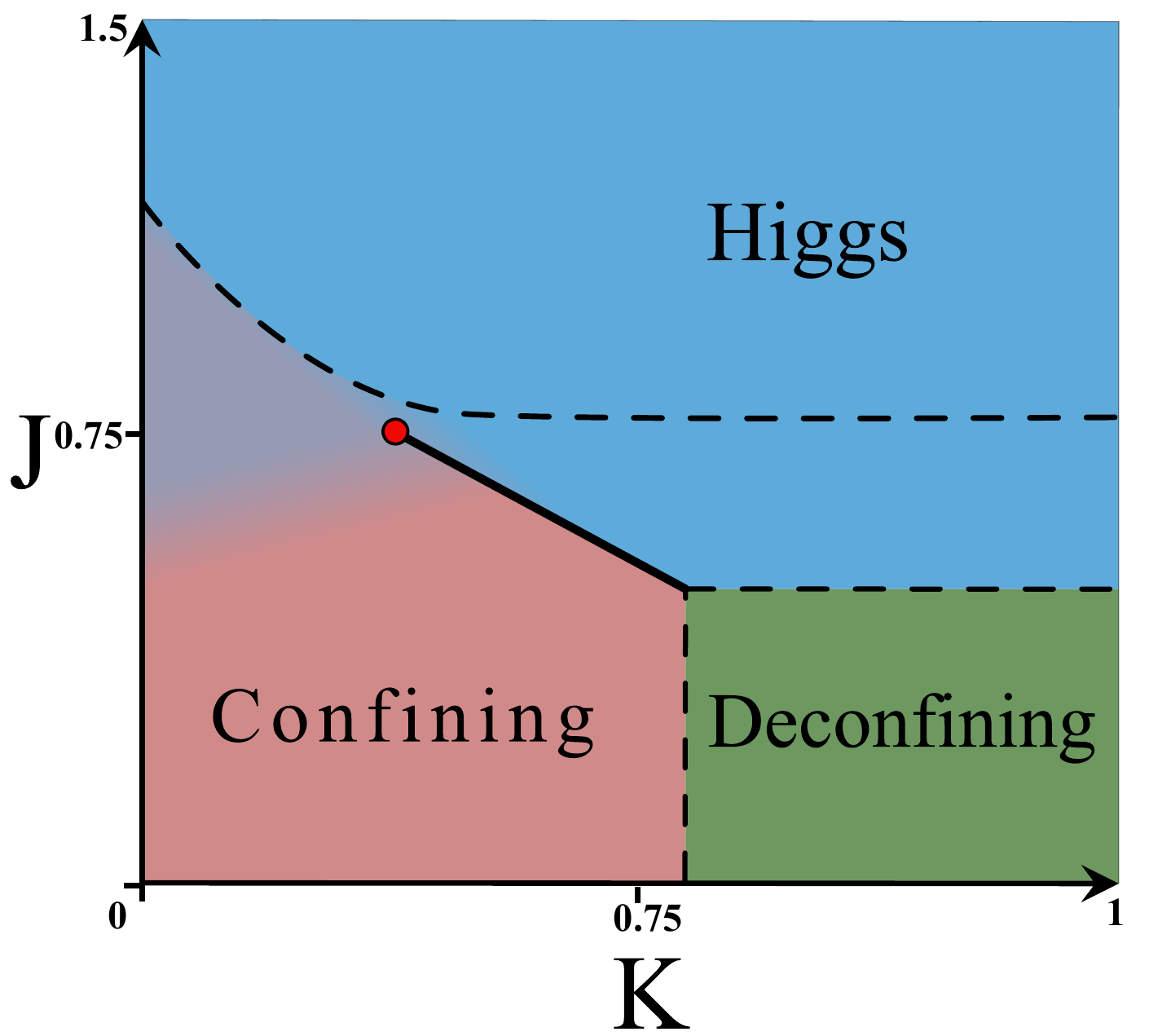}\\
	\raisebox{.3in}{	\includegraphics[width=.28\textwidth]{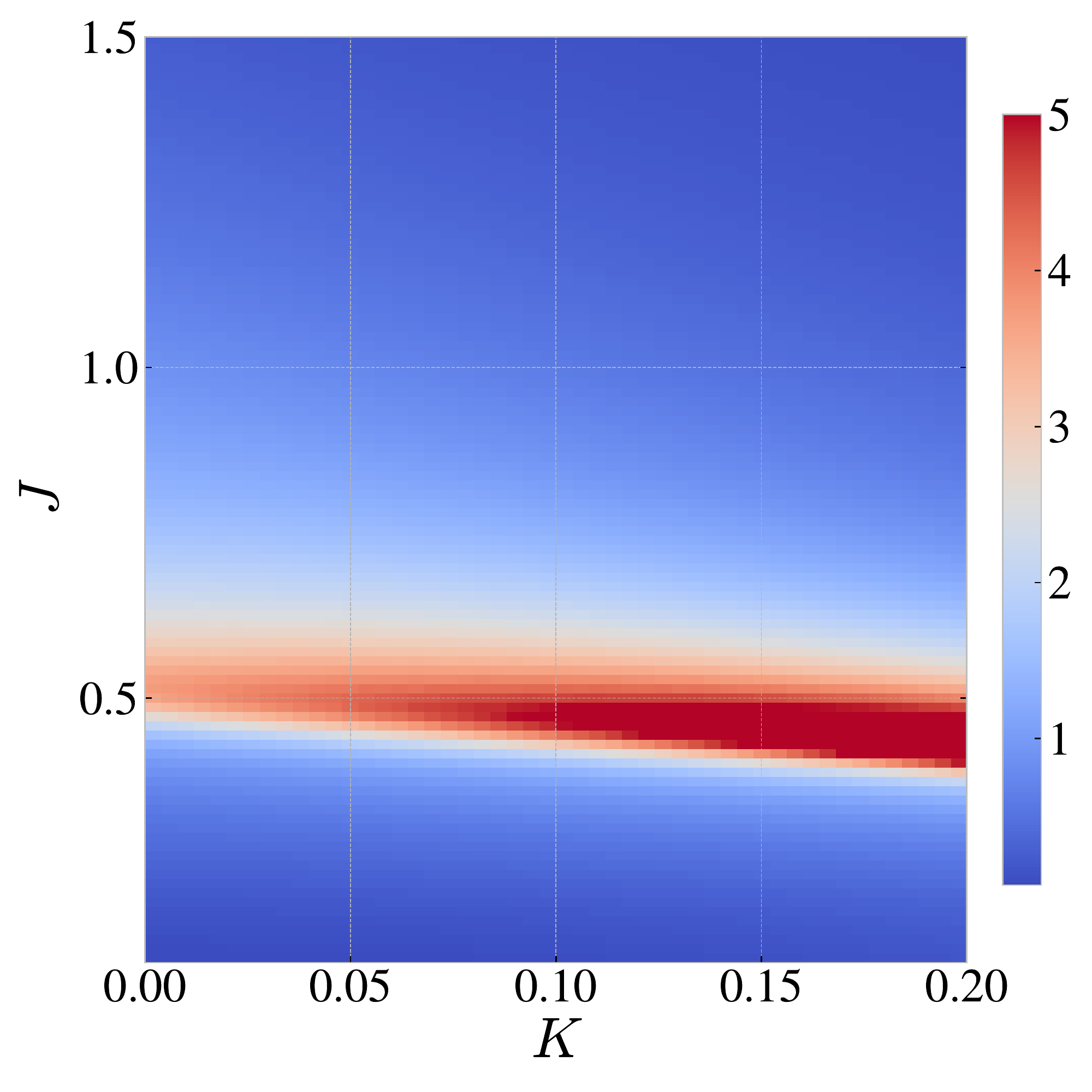}}
	\hfill
        \raisebox{.3in}{
\includegraphics[width=.28\textwidth]{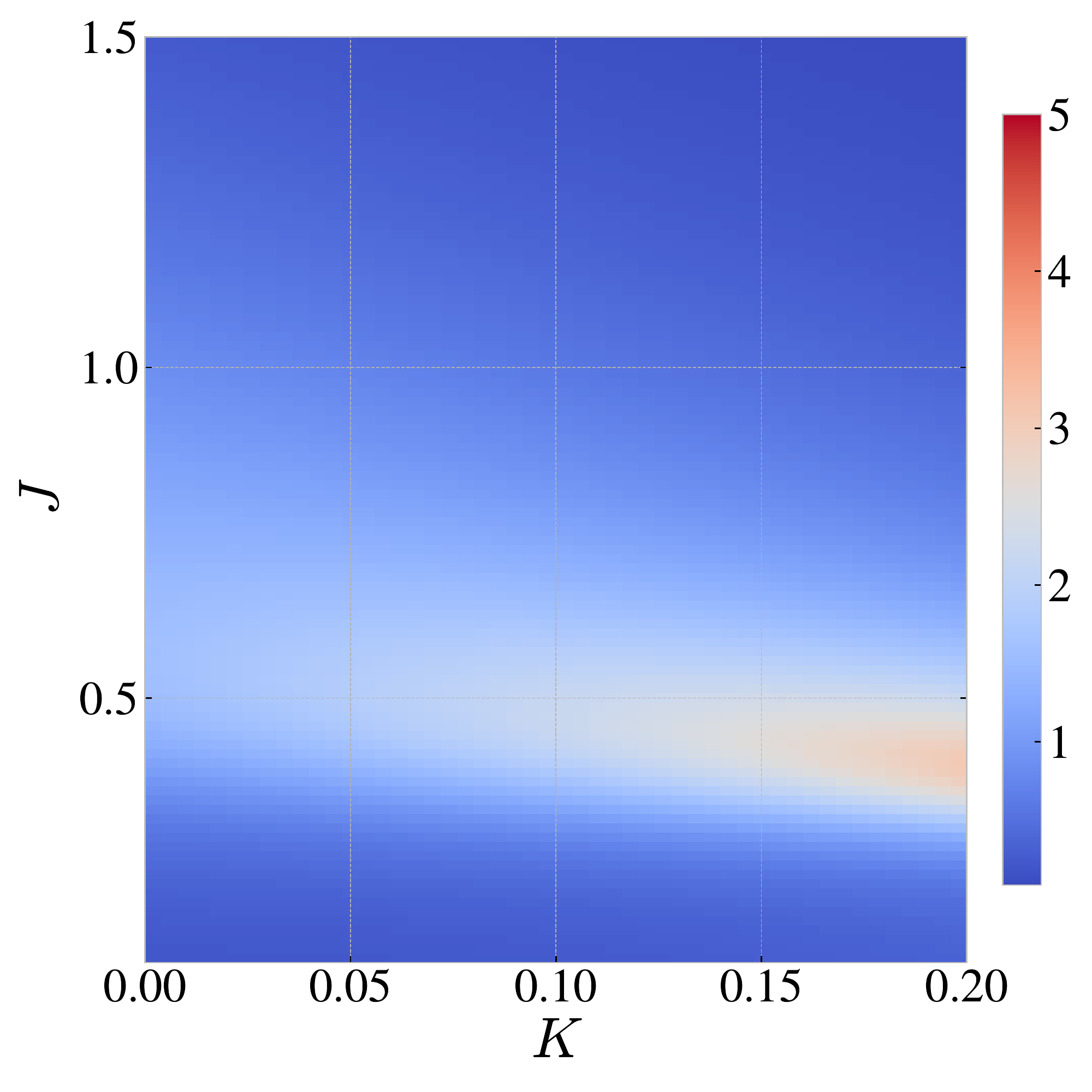}}
	\hfill	\includegraphics[width=.33\textwidth]{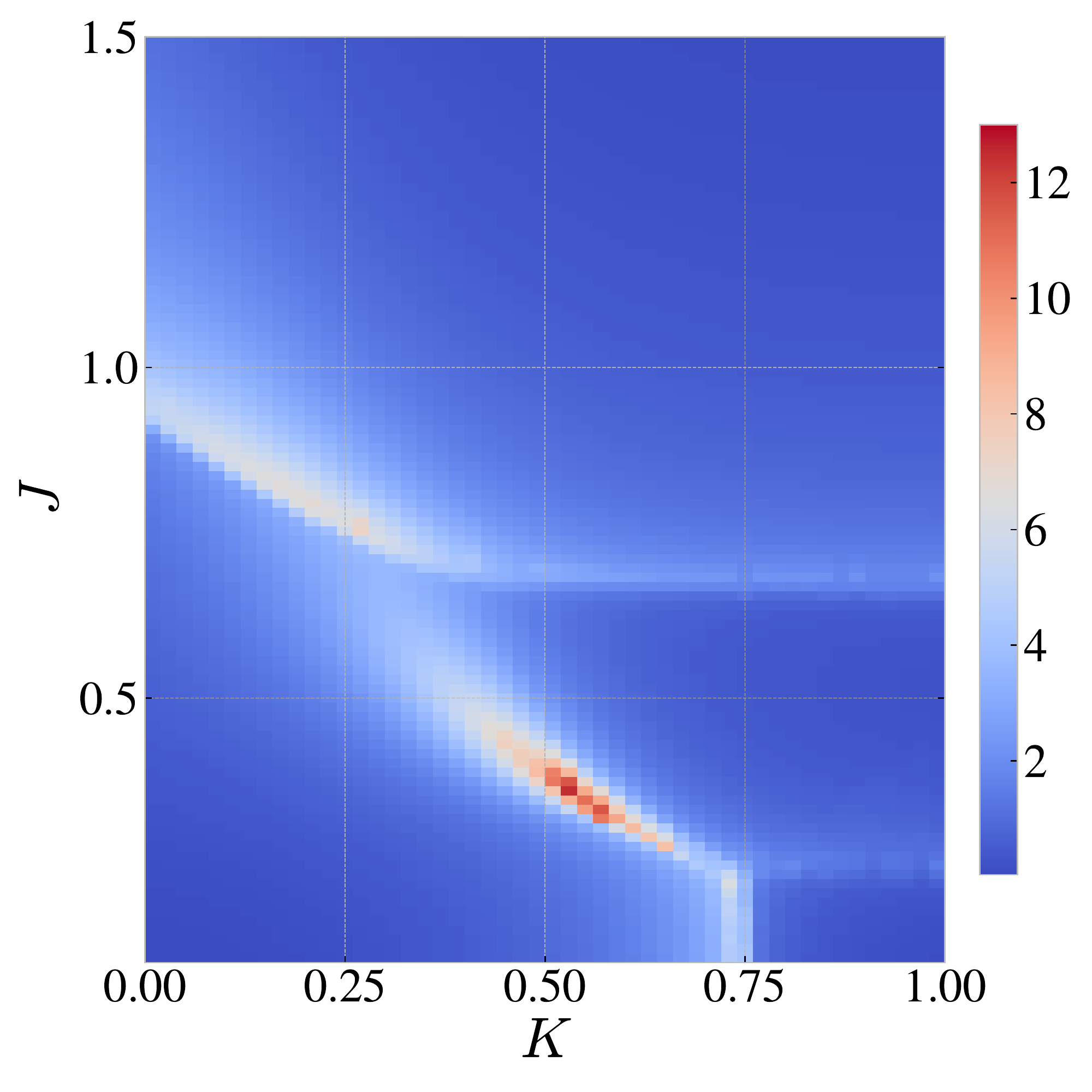}
\\		
	\caption{The phase diagrams of the $Z_2 \cross Z_2 / Z_2$ theory, with the Monte Carlo data for $c_V$ in the second row, for a finite  $J_L=0.1$ (\textbf{a}) and $J_L = 1$ (\textbf{b}). Notice that  the Monte Carlo results are shown over a small interval in $K$ to highlight the most relevant changes.  This local  coupling acts as a field breaking the registry symmetry explicitly, turning the $J_L =0$ Ising phase transition into a cross over. The consequence is that for any finite $J_L$ the indistinguishability of Higgs and confinement is restored. The small strand of second-order like (dashed line) transition for  small $J_L$ (\textbf{a}) may well be related to a rapid crossover that can not be distinguished within our numerical accuracy (Binder criterium) from a real transition. For large $J_L$ (\textbf{b}) the phase diagram becomes nearly identical to the one of the $Z_2/Z_2$ theory itself. Another interesting game is to vary the relative strength of the matter coupling of the two copies for $J_L =0$.  In (\textbf{c}) we show a typical example  dealing with different matter couplings $J = J_1 = 3 J_2$.}
	\label{Z2xZ2Z2Jl}
\end{figure*}

To illustrate these matters further, let us consider what happens when the local $J_L^{12}$ interaction in Eq. (\ref{eq:z2action}) is switched on. It is immediately obvious that this relates directly to the registry: one sees immediately departing from unitary gauge that this lifts the degeneracy of the parallel and anti-parallel registry configurations. This acts as a field breaking the registry Ising symmetry explicitly! This should have the effect to turn the registry phase transition into a cross over, with the effect that yet again the confining and Higgs phases become indistinguishable again, and this is what we find is going on according to our MC simulations: see Fig.\ref{Z2xZ2Z2Jl}  (\textbf{a},\textbf{b}). 

A next freedom one can exploit is to take $J_L =0$ but change instead the relative magnitude of the matter couplings, $J = J_1 \neq J_2$.  This is an entertaining affair highlighting the unusual nature of the registry order. Upon reducing $J_2$ the locus on the vertical axis where this second field changes from order to disorder shifts upwards along the vertical axis in terms of the ``dominating'' copy governed by $J_1=J$. The effect is that at the $J_1$ transition in the small $K$ regime the $J_2$ coupled matter field is still {\em disordered} 
while the order in both fields is required for the existence of the registry order! Hence, holding $J_1\neq J_2$ fixed while varying $K$, there is a window where the registry order disappears in the small $K$ regime, switching on again when $K$ has become sufficiently large to reach the $J$ where also the second field becomes also prone to order again, see Fig.\ref{Z2xZ2Z2Jl}(\textbf{c}). The outcome is a critical end point where now a line of second order transitions starts that is subsequently turned into the first order Huse-Leibler affair! We notice that this peculiar behaviour is only seen in a small $J_2/J_1$ interval; for $J_2 \le 0.25 J$ the phase diagram becomes again the one of the single copy $Z_2/Z_2$ theory.

\bigskip

Up to this point we have demonstrated that in terms of the degrees of freedom of the gauge theory a quantity can be identified characterized by a global symmetry that is broken in the Higgs phase -- the registry order parameter. However, what is the nature of the {\em dynamics} responsible for the stability of this order? Inspecting this deep in the Higgs phase (large $K,J$) employing the unitary gauge is not informative. The reason is that this is rooted in ``gauge field interactions'' that are unusual   in the sense that the discrete nature of the $Z_2$ gauge fields implies that these ``gauge forces'' are characterized by a mass scale that becomes large deep in the Higgs phase. In a statistical physics language any gauge field mediated force may be viewed as an ``order-out-of disorder'' phenomenon -- the fluctuations of the gauge fields are responsible for the interactions between the (gauge invariant) matter fields. 
 
Hence, the limit to consider is $K=0$: the visons are the dynamical degrees of freedom of the discrete gauge theory and these come for free in this limit. Given that the disclinations are bound states of matter defects (domain walls) and the visons, the stability of the Higgs phase itself is entirely due to the cost associated with the former, while the $Z_2$ gauge field is maximally fluctuating. Its only energy cost comes from the coupling to the matter fields. The gauge fields can therefore be straightforwardly integrated out with the outcome that one obtains the (gauge invariant)  Landau-de-Gennes order parameter theory \cite{toner95}. In the general case one is dealing with the point groups associated with the rotational symmetry of the nematic-type state; for Abelian point groups in two dimensions that are encoded by $Z_N$ gauge fields one recovers in this way the simple ``p-adic'' nematics \cite{nematic2D}, while this procedure has been shown to be instrumental to derive the high rank tensor de-Gennes order parameters associated with the non-Abelian point groups in three dimensions \cite{nonabnematic,nonabnematic1,nonabnematic2}. 

The outcome for the elementary $Z_2$ gauge theory is simple: integrating out the gauge fields when $K=0$ leads to the simple ``square of the Hamiltonian'' gauge invariant effective theory as in Eq. (\ref{directorH}) \cite{toner95}. In full generality, departing from the replicated theory with arbitrary matter field symmetry Eq. (\ref{eq:z2action}) when $J_L^{ij} =0$ and all $J_i = J$ one obtains, 
\begin{equation}
H_{\mathrm{eff}, K =0, } = -J' \sum_{\langle i, j \rangle} \left( \sum_{a=1}^{N_\text{rep} } \vec{\phi}_i^a  \cdot \vec{\phi}_j^a \right)^2 ~.
\label{Z2deGennes}
\end{equation}
Let us first consider the single replica $Z_2$ matter field. Here $\vec{\phi_i} \rightarrow \sigma^z_i$ and we infer that the effective Hamiltonian becomes $\sum_{\langle i, j \rangle} (\sigma^z_i \sigma^z_j ) ^2 \rightarrow \mathrm{constant}$: this is the essence of the Fradkin-Shenker observation \cite{Fradkinshenker}, no gauge invariant degree of freedom can be identified distinguishing the Higgs and confining phases implying that these are indistinghuishable. But let us now consider two identical $Z_2$ copies,
\begin{equation}
H_{\mathrm{eff}, K =0, N_{\text{rep}} = 2} = -J' \sum_{\langle i, j \rangle} \left( \sum_{a=1}^{2} (\sigma^z)_i^a  \cdot (\sigma^z)_j^a \right)^2 = 
\mathrm{constant} - J'  \sum_{\langle i, j \rangle} \left( (\sigma^z)_i^{(1)} ( \sigma^z)_i^{(2)} \right) \left( ( \sigma^z)_j^{(1)} (\sigma^z)_j^{(2)} \right)~.
\label{twocopyZ2deGennes}
\end{equation}
This simple affair reveals the origin of the ``registry dynamics'': the combination $\left( (\sigma^z)_i^{(1)} (\sigma^z)_i^{(2)} \right)$ takes the (global) $Z_2$ values $ \pm 1$ for parallel and anti-parallel registry and Eq. (\ref{twocopyZ2deGennes}) is just an Ising Hamiltonian associated with the registry degrees of freedom.
 
In hindsight this is elementary. As for the uniaxial nematics of Toner {\em et al.} \cite{toner95}, the gauge theory in the strong coupling regime ($K \rightarrow 0$) is in fact a redundant parametrization of the ``director'' gauge invariant de Gennes type theories. The additional richness of the gauge theory is associated with $K$ becoming large, i.e. the weak coupling regime of the gauge theory. For instance, the topologically ordered {\em deconfining} phase has no physical identification dealing with the ``molecular'' nematic liquid crystals. Additional microscopic structure is required, with perhaps the ``stripe fractionalization'' \cite{stripefrac,stripefracdemler} being the most elementary example of how this can happen. 
 
This is also underlying the difficulty to recognize this simple  motif ``deep'' in the Higgs phase, for large $J$ and $K \rightarrow \infty$. Departing from the unitary gauge one easily identifies the registry as gauge invariant degree of freedom (as in the above) but at first sight the dynamics stabilizing it is obscure. The reason is of course that the visons are now highly energetic excitations  requiring an energy $E\sim K$ per unit length associated with the effect that the gauge symmetry is discrete and the gauge fields are massive . However, there is no phase transition in the Higgs phase at large $J$, varying $K$ from zero to infinity. This implies that regardless the magnitude of the virtual visons their fluctuations {\em always} suffice to hard wire the registry order. This may be viewed as an order-out-of-disorder phenomenon pushed to its extreme.

\section{The case of many  $Z_2$ matter fields.}
\label{replmanyZ2}

Having established the rules for two $Z_2$ matter fields copies, how does this generalize to many copies? Let us depart again from Eq.(\ref{eq:z2action}) for $Z_2$ matter and consider an arbitrary number of matter field copies $N_{\text{rep}}$. As before the circumstances optimal for the registry type order are associated with setting all local couplings to vanish, $J_L^{ab} = 0$, and taking the matter fields couplings to be equal: $J_a = J$. It is in fact easy to find out by induction how the registry spontaneous symmetry breaking of the two copy case generalizes to many copies.

\begin{figure*}[!h]
	\centering
	\includegraphics[scale=.4]{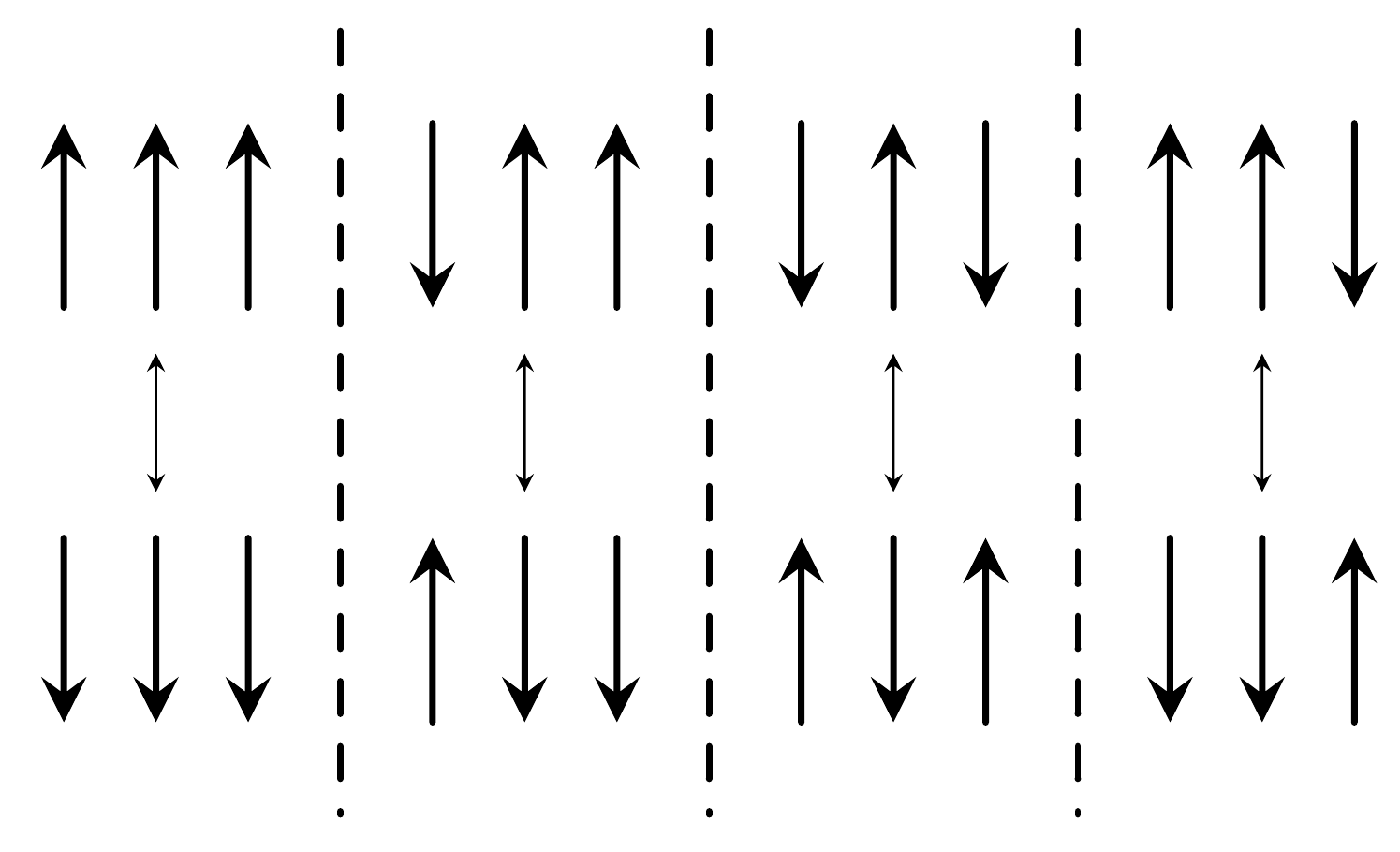}
	\caption{The registry order parameter (see main text) in the case of the three copy $Z_2 \times Z_2 \times Z_2/Z_2$ theory. Depart from unitary gauge fix and the three matter fields can locally form eight configurations. However, under gauge transformations half of them are redundant, e.g. $\uparrow \uparrow \uparrow \leftrightarrow \downarrow \downarrow \downarrow$. The gauge invariant registry order parameter takes therefore four different physical realizations and it is governed by a $p=4$ state Potts model.}
	\label{threecopyexample}
\end{figure*}

Let us consider the three copy case:  $Z_2 \times Z_2 \times Z_2/ Z_2$. We just proceed as before, departing from deep in the Higgs phase (large $K,J$) and using the unitary gauge. At every site the matter fields can occur in $2^3$ different configurations, see  Fig.\ref{threecopyexample}. Identifying these with domains one would find accordingly 8 different domain walls associated with flipping one spin keeping the other spins fixed. However, upon restoring the gauge invariance amounting to flipping {\em all} spins on the site, it follows that configurations are pairwise associated, e.g.  $\uparrow \uparrow \uparrow \leftrightarrow  \downarrow \downarrow \downarrow$. Accordingly, one finds 4 distinct gauge invariant vacuum states, separated by ``single spin'' domain walls. This is governed by a $p = 4$ state {\em Potts model}! It is easy to check that the Ising registry theory Eq.(\ref{twocopyZ2deGennes}) generalizes to the 4-state Potts theory  by considering the ``squared'' de-Gennes type effective theory in the $K \rightarrow 0$ limit. 

\begin{figure*}[!h]
	\centering	\includegraphics[width=0.425\textwidth]{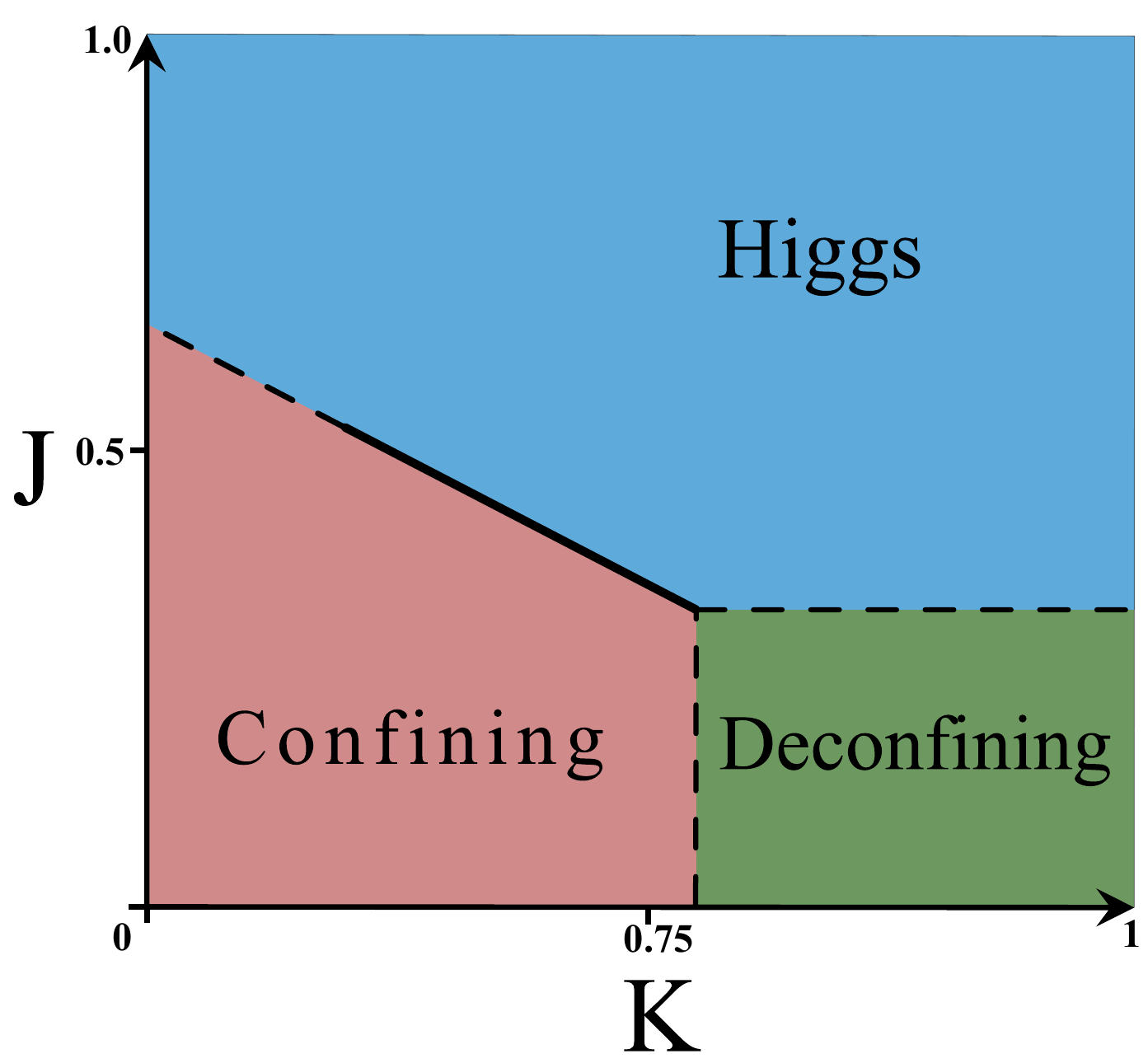}
        \raisebox{-.15in}{
	\includegraphics[width=0.45\textwidth]{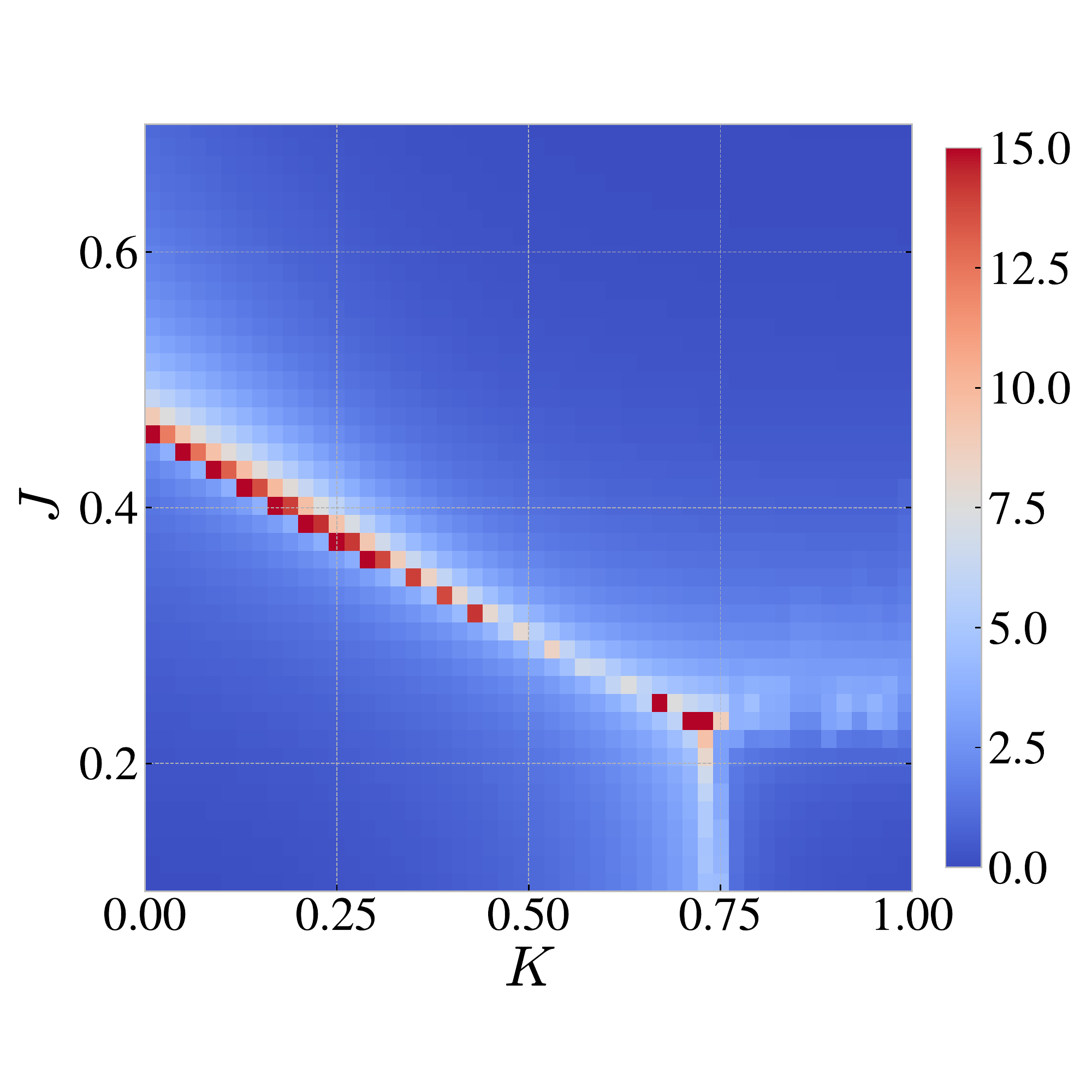}}
	\caption{The phase diagram of the three copy $Z_2 \cross Z_2  \cross Z_2/ Z_2$ theory as in Fig.\ref{Z2Z2phasediagram} and Fig.\ref{O2Z2phasediagram} for the ``minimal'' $J_a =J$ and $J^L_{ab} = 0 $ case. This looks very similar to Fig.\ref{Z2xZ2Z2phasediagram} and the strand of the first order transitions is barely changed. The specific heat $c_V$ data suggests a first order transition all the way down to $K=0$, but a Binder cumulant study reveals this change to second order as sketched. A major distinction now is that dynamics of (continuous) confining to Higgs transition is governed by a $p=4$ state \textit{Potts model}.} 
	\label{Z2xZ2xZ2Z2phasediagram}
\end{figure*} 

\begin{figure*}[!h]
	\centering
	\includegraphics[scale=1]{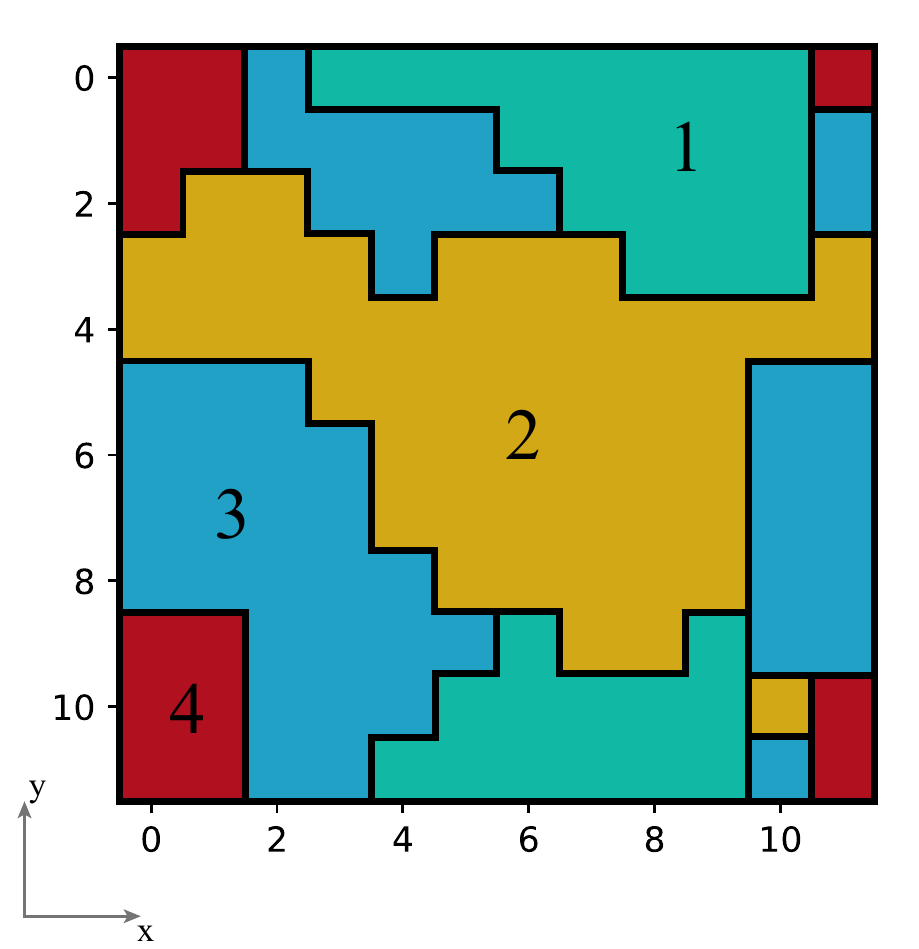}
	\caption{Planar snapshot along the $z$ axis of a Monte Carlo quench for the three copy model deep in the Higgs phase $J_1=J_2=J_3=0.8$ and $K=0$. This reveals the presence of the 4-state Potts model domains of Fig.\ref{threecopyexample} separated by the ``registry'' domain walls. Domain 1 - \fcolorbox{black}{z1}{\rule{0pt}{6pt}\rule{6pt}{0pt}} are $\uparrow \uparrow \uparrow \leftrightarrow \downarrow \downarrow \downarrow$, domain 2 - \fcolorbox{black}{z2}{\rule{0pt}{6pt}\rule{6pt}{0pt}} are $\downarrow \uparrow \uparrow \leftrightarrow \uparrow \downarrow \downarrow$, domain 3 - \fcolorbox{black}{z3}{\rule{0pt}{6pt}\rule{6pt}{0pt}} are $\downarrow \uparrow \downarrow \leftrightarrow \uparrow \downarrow \uparrow$, domain 4 - \fcolorbox{black}{z4}{\rule{0pt}{6pt}\rule{6pt}{0pt}} are $\uparrow \uparrow \downarrow \leftrightarrow \downarrow \downarrow \uparrow$. Every domain wall differs from the neighboring domain by a single flip.}
	\label{Z2xZ2xZ2Z2snapshot}
\end{figure*}

This is confirmed by our MC simulations. In Fig.\ref{Z2xZ2xZ2Z2phasediagram} we show the phase diagram of the three copy model. As anticipated, this looks very similar as to the two  copy case, Fig.\ref{Z2xZ2Z2phasediagram}. Although the dynamics is quite different --- the 4 distinct ``gauged'' domain walls associated with $p=4$ Potts --- the Huse-Leibler first order strand is barely affected with yet again the registry order being responsible for the continuous (Potts model) phase transition distinguishing the confining and Higgs phases. In Fig.\ref{Z2xZ2xZ2Z2snapshot} we show a typical  realization after a partially annealed Monte-Carlo ``quench'' deep in the Higgs phase: one discerns the 4 distinct Potts domains separated by the ``registry domain walls'' confirming this simple analysis.  

As for the two copy case one can now proceed by switching on  various $J_L^{ab}$ local couplings, ``gluing together''  the local matter copies with the expected results that we checked. Involving one coupling between two of the three sectors, the 4-state Pott symmetry is lifted by the explicit symmetry breaking to the effective two copy registry Ising symmetry. Similarly, coupling all copies with each other diminishes the registry spontaneous symmetry breaking such that confinement and Higgs become indistinguishable. In the same guise one can detune the matter couplings $J_a$, turning into a variation of the matters we discussed in the previous section. 
 
The two and three copy cases reveal the counting rules and by induction we can generalize this now to an arbitrary number copies. We just proceed as for the three copy case. Given $N_{\text{rep}}$ copies there are a total of $2^{N_{\text{rep}}}$ local configurations in unitary gauge. The next observation is that these configurations are pair wise identified with each other by the gauge transformation. The result is that the registry symmetry is now captured by a $p = 2^{N_{\text{rep}}} /2 $ state Potts model. 

\section{Replicating $O(2)$ matter.}
\label{replO2}

\begin{figure*}[!h]
  \centering	\includegraphics[width=0.425\textwidth]{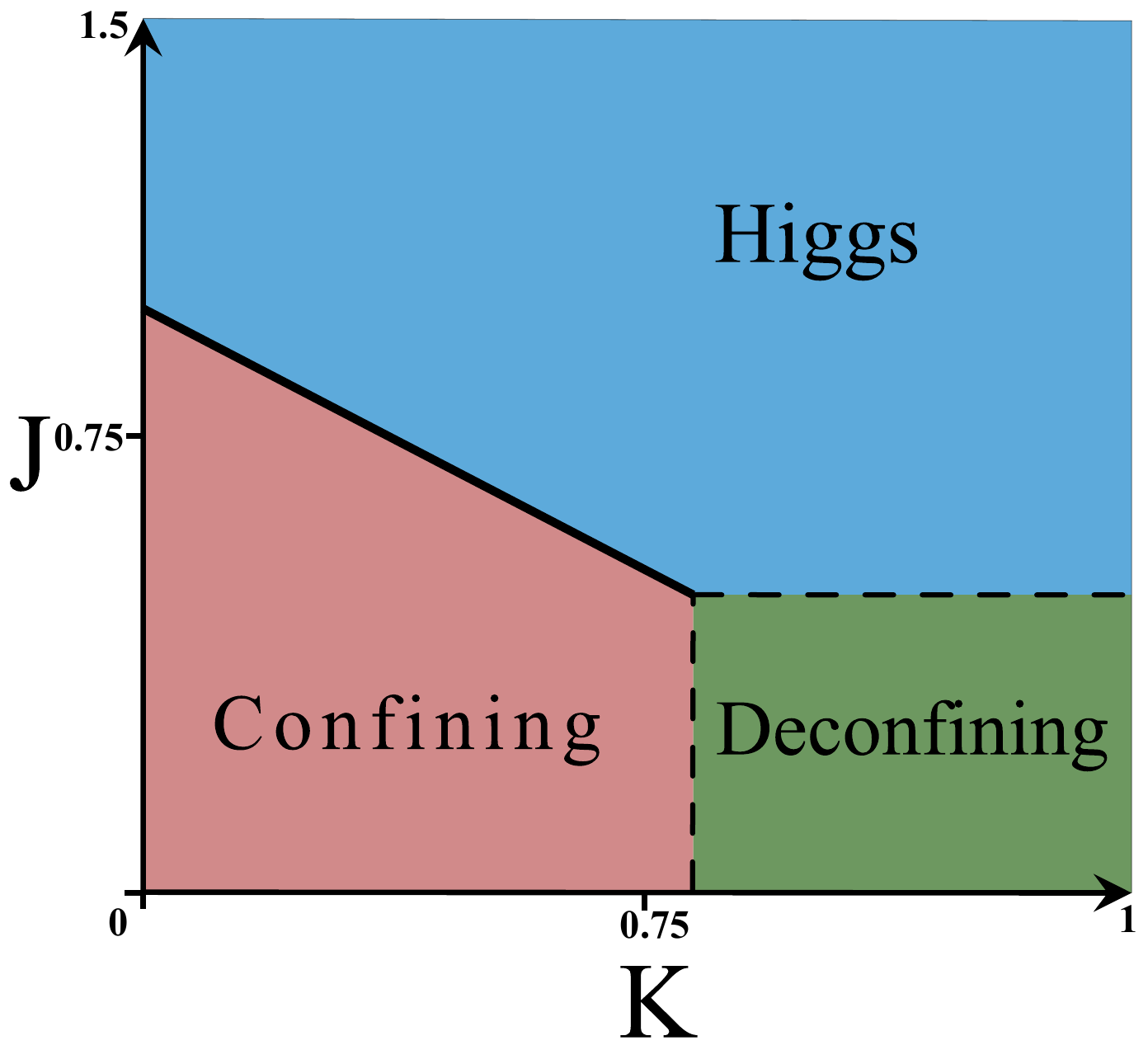}
  \raisebox{-.15in}{
	\includegraphics[width=0.45\textwidth]{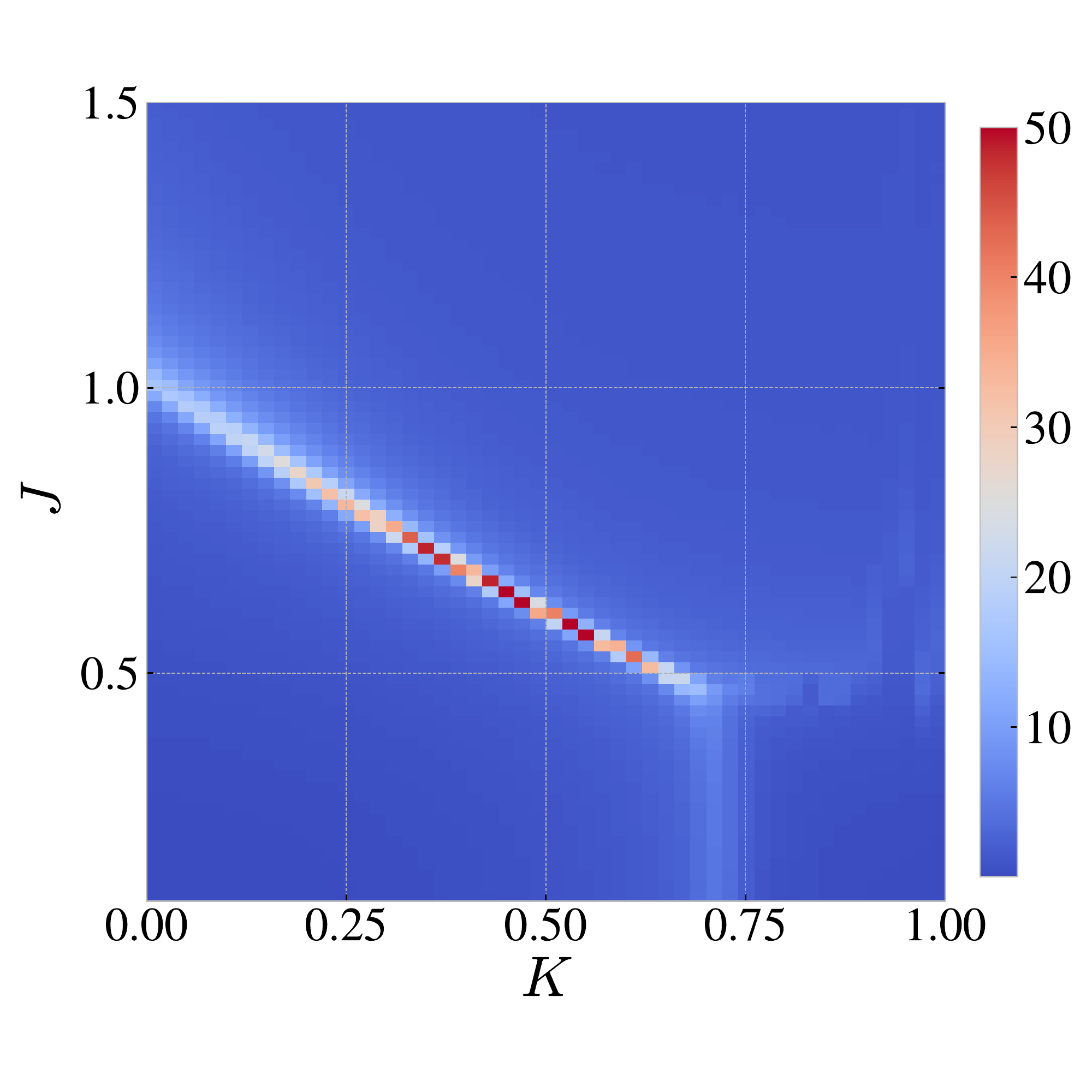}}
	\caption{The phase diagram of the $O(2) \cross O(2) / Z_2$ (left) and raw MC data for $c_V$(right) for $J_1 = J_2 = J$ and $J_L^{12} =0$. This is yet again remarkably similar as to the other phase diagrams. The difference is that the Higgs-confinement phase transition is now first order all the way to $K=0$ contrasting with its second order behavior for the single copy version, Fig.\ref{O2Z2phasediagram}. This due to the occurrence of a Ising type registry order parameter in the Higgs phase which is of the same kind as for the $Z_2$ matter cases, that disappears simultaneously with the nematic type order parameter upon entering the confining phase. The first order nature is confirmed by a Binder cumulant study (Fig.\ref{weakfirstorder}{\bf (c)} in Appendix \ref{binder}).}
	\label{O2xO2Z2phasediagram}
\end{figure*}

The ``matter in the fundamental'' $Z_2$ matter case is special, and what to expect when the symmetry of the matter field is raised relative to the $Z_2$ gauge 
symmetry? For this purpose we focused in on the minimal extension: the replicated $O(2)$ matter fields gauged by $Z_2$. As we discussed in Section \ref{Z2Z2review}, for a single copy the Higgs phase is now characterized by the nematic-like order, distinguishing it from the confining phase through the presence of a second order phase transition.
 
In Fig.\ref{O2xO2Z2phasediagram}  we show the phase diagram of the two copy $O(2)\cross O(2)/Z_2$ case for the (usual) choice $J=J_1=J_2, J_L^{12} = 0$. This looks yet again very similar as to the other cases, the main difference being that now the transition between the Higgs and confining phase has turned into a {\em first order} transition for {\em all} $K$. Inspecting the ``strength'' of the first order transition exploiting the Binder criterium (see Appendix \ref{binder}) we find that close to the tricritical point the transition looks quite like the other cases: this is clearly driven by the ``Huse-Leibler'' amplitude fluctuations. Upon reducing $K$ the transition becomes an increasingly weak first order transition, but it continues to be first order all the way down to $K \rightarrow 0$.
 
As we will argue, this first order behavior is due to the fact that {\em two order parameters governed by two independent global symmetries vanish simultaneously}  at the Higgs to confinement transition. In fact, the replicating has the effect that the ``accidental'' second order nature of this transition for $O(2)/Z_2$ becomes similar to the generic first order transition of the $O(N)/Z_2$ system with $N \ge 3$ as argued by Ref. \cite{toner95}.

 What are the two symmetries that are broken in the Higgs phase?  This is actually a bit more of a subtle affair than for the simple $Z_2$-replica's. As we will argue and confirm with the Monte Carlo, one type of symmetry is associated with the nematic type ``halved periodicity'' XY as for a single $O(2)/Z_2$, actually applying to both copies individually. But these are non-locally coupled together by the gauge fluctuations in a way that they submit to a perfect $Z_2$ registry symmetry, that is macroscopically the same symmetry as the registry $Z_2$ of the $Z_2\times Z_2/Z_2$ theory.        
 
This is yet again easy to deduce by zooming in on the maximal gauge fluctuations, the $K=0$ case. As discussed in Section \ref{replZ2}, this is of the universal form Eq.(\ref{Z2deGennes}). For the two copy $O(2) \times O(2)/Z_2$  case it follows immediately,
\begin{eqnarray}
	H_{O(2) \times O(2), K \rightarrow 0} & =  & -J' \sum_{\langle i, j \rangle} \left( \vec{\Phi}_{i1} \cdot  \vec{\Phi}_{j1} + \vec{\Phi}_{i2} \cdot  \vec{\Phi}_{j2} \right)^2 \nonumber\\
	& \sim & -J' \sum_{\langle i, j \rangle}  \left( ( \vec{\Phi}_{i1} \cdot  \vec{\Phi}_{j1})^2  + (\vec{\Phi}_{i2} \cdot  \vec{\Phi}_{j2})^2 +2 (  \vec{\Phi}_{i1} \cdot  \vec{\Phi}_{j1} ) \times ( \vec{\Phi}_{i2} \cdot  \vec{\Phi}_{j2}   ) \right) \mathperiod  
	\label{O2O2effH}
\end{eqnarray}
The first and second term just represent the director order parameter -- for $O(2)$ just the halved periodicity -- and upon ignoring the last term one is just dealing with two completely decoupled identical "$O(2)$ nematics".  This last term encapsulates the interactions between these two copies as induced by the fluctuating disclinations. It is obvious that also this term is governed by an invariance under $O(2)$ rotations of every copy separately -- there is surely no effective single site "anisotropy" at work reducing  it to the on site  $Z_2$ registry revealed by Eq. (\ref{twocopyZ2deGennes}) of the $Z_2 \times Z_2 / Z_2$ case.

\begin{figure*}[!h]
	\centering
	\includegraphics[scale=.5]{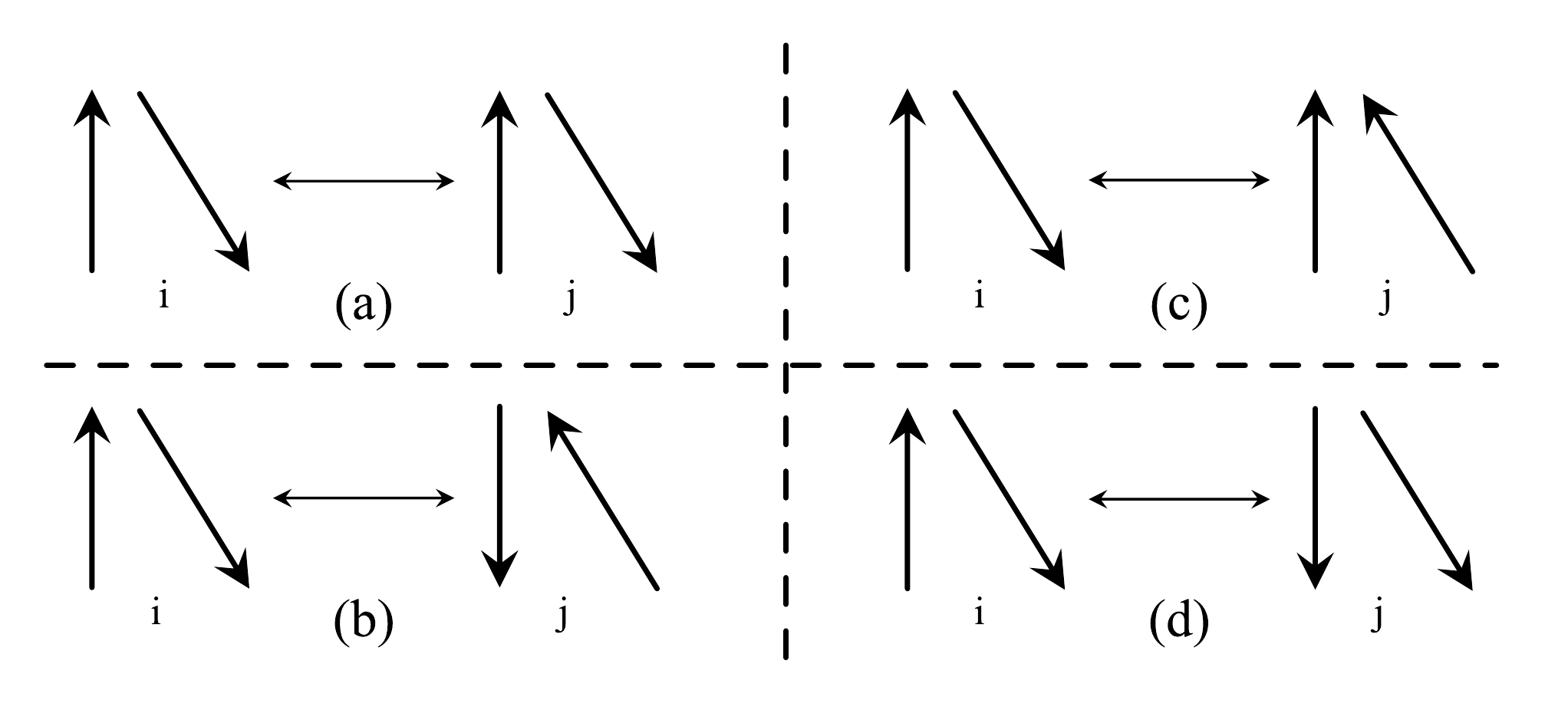}
	\caption{Illustration of the registry order of the  $O(2)\cross O(2) / Z_2$ theory: the construction of the registry Ising domain wall. We see that in contrast with the multiple $Z_2$ matter fields --- the \textit{local onsite} orientation of different matter fields is irrelevant for the registry. The two copies acquire independent orientations on the $O(2)$ (half) circle.   
          What matters is how two matter fields change {\em together} from site to site. The strong gauge field fluctuations ``glue'' together matter fields in a sense that they would have to simultaneously change together in the same direction, see Eqs.\eqref{O2O2effH},\eqref{O2O2dyn}.  In all cases \textbf{(a)}-\textbf{(d)} we depart from the same unitary gauge two spin reference configuration on site $i$. In \textbf{(a)} we just consider the ground state configuration on site $j$ in unitary gauge. By performing a gauge transforming on site $j$ we obtain  \textbf{(b)}, gauge equivalent  to \textbf{(a)}. But now consider  \textbf{(c)}, with an anti-parallel orientation of the second spin on site $j$, 
          gauge equivalent with \textbf{(d)}. 
The case 	\textbf{(a)}, \textbf{(b)} corresponds with $\nabla_{ij} \phi_1 = \nabla_{ij} \phi_2=0$. We see that these configurations minimize the interaction energy Eq.\eqref{O2O2dyn}.  While \textbf{(c)}, \textbf{(d)} instead has $\nabla_{ij} \phi_1 =  0, \nabla_{ij} \phi_2=\pi$. Compared to \text{\bf(a)}-\textbf{(b)} this would cost us $2J^{\prime}$ energy. The configuration \textbf{(a)}-\textbf{(b)} corresponds with a registry domain wall  between \textit{parallel} and \textit{anti-parallel} orientations of $O(2)$ matter directors at the neighboring sites.}
\label{O2O2Z2order}
\end{figure*}

However, the registry is now hidden in the "synchronization" imposed by the gauge fluctuations associated with the relative orientation of the two spin configurations on neighbouring sites.  Parameterize $\vec{\Phi} = | \Phi | e^{i \phi}$, such that $\phi_{j a} = \phi_{i a} + \nabla_{ij} \phi_a$ and the interaction term becomes,
\begin{equation}
	-J' \sum_{\langle i, j \rangle} (  \vec{\Phi}_{i1} \cdot  \vec{\Phi}_{j1} ) \times ( \vec{\Phi}_{i2} \cdot  \vec{\Phi}_{j2}   ) =   -J' | \Phi |^2 \sum_{\langle i, j \rangle} \cos ( \nabla_{ij} \phi_1 ) \cos(  \nabla_{ij} \phi_2) \mathperiod
	\label{O2O2dyn}
\end{equation}
This interaction term is governed by a global $Z_2$ symmetry! It originates in the synchrony of gauge invariant ``remnants'' of the two copies: the interaction term is minimal either for {\em both} copies being parallel on neighbouring sites  or {\em both} antiparallel where $\nabla_{ij} \phi_a=0$ or $\nabla_{ij} \phi_a=\pi$.  This signals the two fold degeneracy of the Ising order. In Fig.\ref{O2O2Z2order} we illustrate how to  construct the Ising domain walls associated with this registry order.  

\begin{figure*}[!h]
	\centering
	\includegraphics[width=0.45\linewidth]{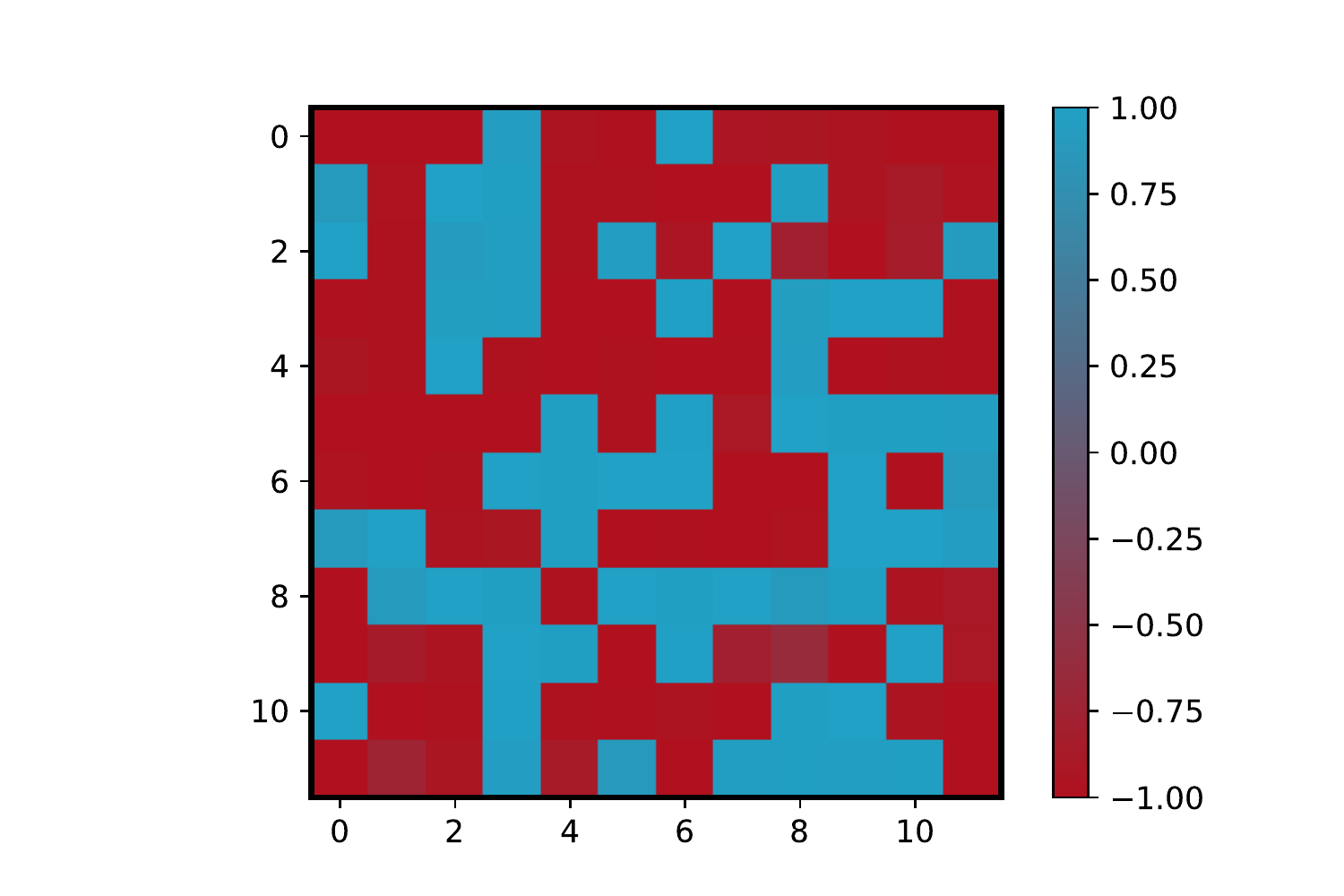}
	\includegraphics[width=0.45\linewidth]{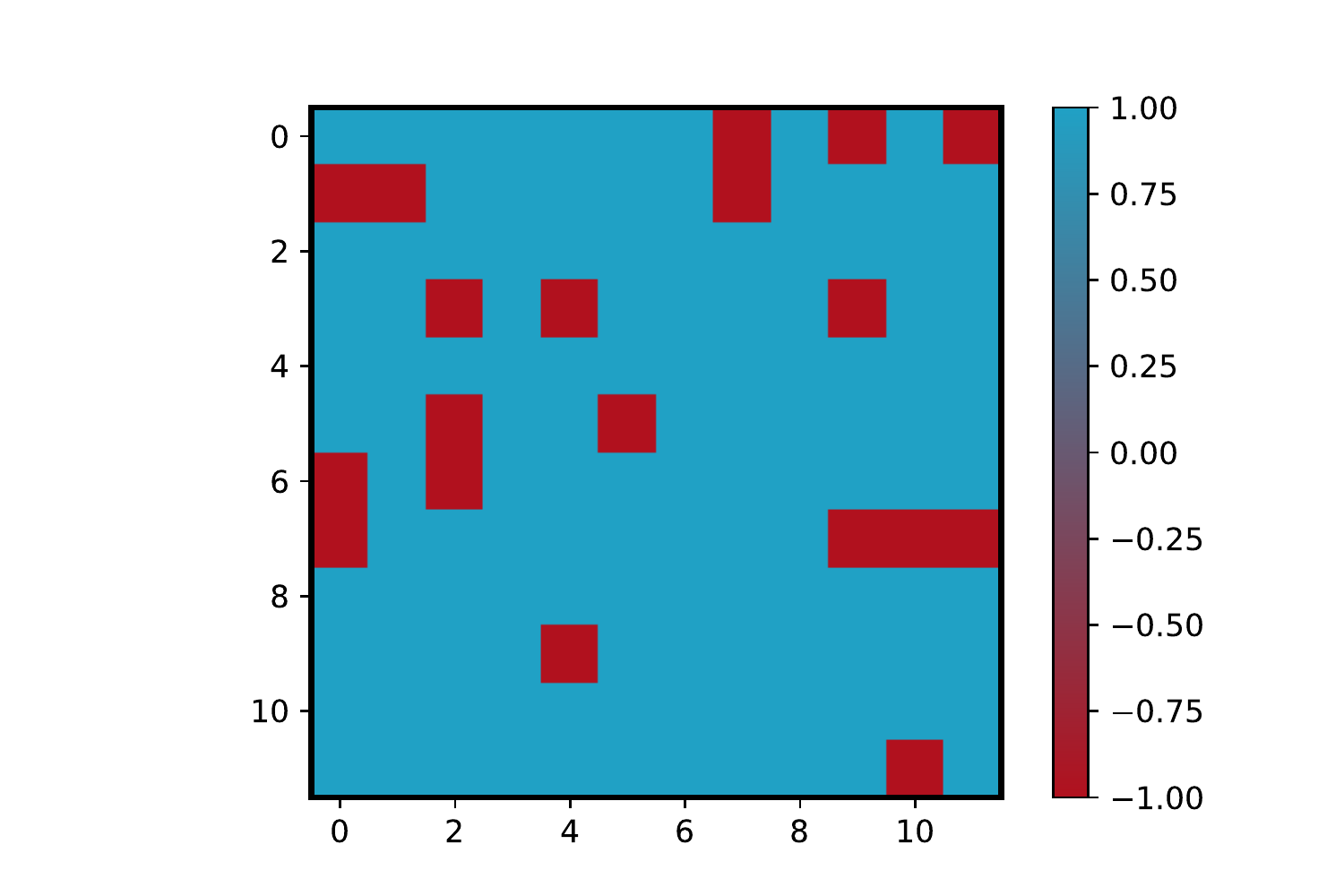}
	\caption{2D snapshot along $z$ axis of a system on $12\cross12\cross12$ lattice with periodic b.c for $O(2) \times O(2) / Z_2$  model. We a single frozen configuration of fields computed using a Monte Carlo quench deep in the Higgs phase: $J_1=J_2=5$ and $K=0$. \textbf{(Left)} Snapshot of $\cos(\nabla_{ij}\phi_1)$. This elucidates a simple registry order (parallel/anti-parallel), in the form of registry domains that are frozen in. The color coding denotes the cosine of the relative orientation of the spins on the neighboring sites: Domain 1 (parallel) - \fcolorbox{black}{z3}{\rule{0pt}{6pt}\rule{6pt}{0pt}} corresponds with $\uparrow \uparrow \leftrightarrow \downarrow \downarrow$, and domain 2 (anti-parallel) - \fcolorbox{black}{z4}{\rule{0pt}{6pt}\rule{6pt}{0pt}} accordingly  $\downarrow \uparrow \leftrightarrow \uparrow \downarrow$. \textbf{(Right)} Snapshot of $\exp(2i\phi_1)$ director order parameter showing homogeneous distributions, being annealed, with some local fluctuations due to the finite size and temperature effects.}
	\label{O2O2Z2snapshot}
\end{figure*}

This can be illustrated by a Monte-Carlo quench from ``high temperature'', departing from random configurations and partially annealing the system in the Higgs regime, similar to Fig.\ref{Z2xZ2xZ2Z2snapshot}. This is shown in Fig.\ref{O2O2Z2snapshot}. One infers that one form of order is of the nematic-type (XY like)  associated with the orientation of the director of one of the the copies that we find to be homogeneous in this snapshot -- this is annealed. However we can also track the {\em relative orientation of a single field between neighboring sites} $\nabla_{ij}\phi_\alpha$, for $\alpha=1$. Fig.\ref{O2O2Z2snapshot} (Left) shows that this is either parallel $\nabla_{ij}\phi_1=0$ or anti-parallel $\nabla_{ij}\phi_1=\pi$ as illustrated in Fig.\ref{O2O2Z2order} and explained after the text. As in Fig.\ref{Z2xZ2xZ2Z2snapshot} we observe domains of the two different parallel and anti-parallel registry separated by ``registry domain walls'' of the kind similar to the $Z_2$ matter.
 
Similar to the $Z_2\cross Z_2/Z_2$ case, this registry order is critically dependent on the $J_L^{12}$ being zero -- upon switching it on this acts as an explicit symmetry breaking of the registry turning the first order transition near $K=0$ into second order, exhibiting a phase diagram that is like the single  copy $O(2)/Z_2$ case (Fig.\ref{O2Z2phasediagram}). The effect of unbalancing the matter couplings ($J_1 \neq J_2$) has very similar effects as illustrated in Fig.\ref{Z2xZ2Z2Jl}, although now a second order transition is left behind when the registry order switches off.

 In summary, the ``registry-sector'' of the $O(2)$ case behaves more or less identically as in the $Z_2$ case. The difference is that in the $O(2)$ case one also has to account simultaneously for the nematic type order characterizing this Higgs phase. This renders the phase transition first order all the way to $K=0$. 
The topological defects of this nematic-like state will start to populate the vacuum upon lowering $K,J$. These are the disclinations, in turn being a ``confined'' combination of the vortex-type matter defect, and the $Z_2$ gauge flux/vison, e.g. \cite{toner95,Senthil}. The matter-vortices of the two copies share a single gauge flux. Upon integrating out these ``topological fluctuations'', an interaction mediated by the $Z_2$ gauge fields develops which is responsible for the registry order.   Similar to the $Z_2$ case, the limit where one can easily deduce the effects of the visons ``gluing'' the copies into registry, this is most easily deduced in the $K=0$ limit which is entirely controlled by the matter interaction $J$.    

\section{Discussion and conclusions}
\label{discussion}

Gauge field theory is of course well known to have its own rules. In this paper we have focused in on the simplest of all gauge symmetries, the $Z_2$ variety, as the simplest theory revealing the characteristic phase structure characterized by confinement, deconfinement and the Higgs phase. By introducing the matter replicas we discovered a new set of phenomena. In this pursuit we have heavily leaned on the unbiased Monte-Carlo simulations. Puzzled by the outcomes we discovered the new phenomenon of ``registry order parameter''. As we discussed in Sections \ref{replZ2} and \ref{replmanyZ2}, it is very easy to identify the origin in the $Z_2$ matter versions, although it is a bit less obvious and arguably more entertaining  for the $O(2)$ version in Section \ref{replO2}. It took us by surprise, given that the origin of the induced ``gauge interaction'' that is responsible for the registry symmetry breaking  is of a kind that is rather unfamiliar.  
 
The mechanism is revealed by considering the extreme strongly coupled limit of the gauge fields, $K=0$. In terms of the degrees of freedom of the gauge theory, the mechanism is unusual, highlighted by the $O(2)$ case. The physical degrees of freedom associated with the disordering of the Higgs phase are the nematic-type disclinations that are in turn confined combinations of matter vortices and the fluxes of the gauge field (the visons). Although both matter fields carry their own vortices, these ``share'' a single vison. In the $K=0$ limit the latter come for free and upon integrating these out one finds that the de-Gennes type effective order parameter theory is endowed with the registry ``Ising'' Hamiltonian, Eq . (\ref{twocopyZ2deGennes}), and Eq.(\ref{O2O2dyn}) respectively. Interestingly, the discrete nature of the gauge theory has eventually the effect to generate the Ising type registry global symmetry breaking. 
  
We have only inspected the most elementary forms of such gauge theoretical systems. 
These are just a point in the vast landscape of all gauge theories, up to the non-Abelian Yang-Mills theory behind e.g. the Standard Model. It would be quite interesting to find out what happens with this ``registry order'' upon systematically raising the symmetries involved. What happens in the ``replicated''  $Z_2$ gauge theory involving the non-Abelian $O(N)$ matter fields with $N \ge 3$? What happens raising the $Z_2$ gauge symmetry to the non-Abelian 3D point group symmetries as in Ref. \cite{nonabnematic}? Even more fundamental, what is the fate of registry dealing with {\em continuous} gauge symmetry, starting with the Abelian $U(1)$ of compact electrodynamics \cite{PolyakovQED}? 
 
Finally, another aspect also caught us by surprise in the Monte Carlo outcomes for the various phase diagrams. The Huse-Leibler mechanism for the first order transition between the "Higgs'' and "confinement-like phase'' emanating from the tricritical point appears to be surprisingly universal. Eventually this is a quantitative affair. The mechanism as understood for the $Z_2/Z_2$ case does in this regard rely on the specifics of this particular theory: the ``vesicles'' versus ``platelet'' affair. This surely works differently involving continuous symmetry -- the $O(2)$ cases. But surprisingly the ``first order strand'' is even quantitatively barely affected by these fundamental differences. The reason for this is presently unclear to us and it may be of interest to have a closer look at the origin of this ``quasi-universality''.

\acknowledgments
We thank T. Senthil and K. Liu for discussions. 
This research was supported in part by the Dutch Research Council (NWO) project 680-91-116 ({\em Planckian Dissipation and Quantum Thermalisation: From Black Hole Answers to Strange Metal Questions.}), and by the Dutch Research Council/Ministry of Education. The numerical computations were carried out on the Dutch national Cartesius and Snellius national supercomputing facilities with the support of the SURF Cooperative as well as on the ALICE-cluster of Leiden University. We are grateful for their help.

\appendix

\section{Appendix}
\subsection{Monte Carlo Simulations}
\label{app:mc-details}
The Monte Carlo simulations were performed on a $d \times d \times d$ grid with periodic boundary conditions. Grid size in all simulations was $d=12$ in order to avoid finite size effects and still get reasonable computational times. We used most of the time a number of  measurement sweeps $N=6000000$; thermalization sets in typically after 1/3 of the sweeps. Near the critical points we checked this by tracking the evolution of the various quantities  as function of the number of steps, taking as many steps as needed for the quantity to saturate. For the updating rules we used the  Metropolis-Hastings algorithm with the acceptance ratio $A(n, n^\prime) = \min(1, e^{-\Delta E_{n,n^\prime}})$ where $\Delta E_{n,n^\prime}$ is the energy difference between states $n$ and $n^\prime$ that differ in a single matter field or gauge field flip. Phase diagrams were obtained by vertically scanning along different values of $J$'s, using annealing order to improve convergence accuracy. The whole phase diagram was run on a remote cluster where each process was  associated with a single value of $K$ scanning along the $J$ axis. We noted that the longest sweeps were required to equilibrate when deep in the deconfining regime. Flipping single ``bond spins'' using the Metropolis-Hastings algorithm leads to highly energetic configurations and accordingly to long thermalization times. But this did not pose any difficulty since the physics in this regime is simple.

\subsection{Determining order of phase transition}
\label{binder}
\begin{figure*}[!h]
  \centering
  (\textbf{a})
  \raisebox{-1.1in}{
 \includegraphics[width=.2\textwidth]{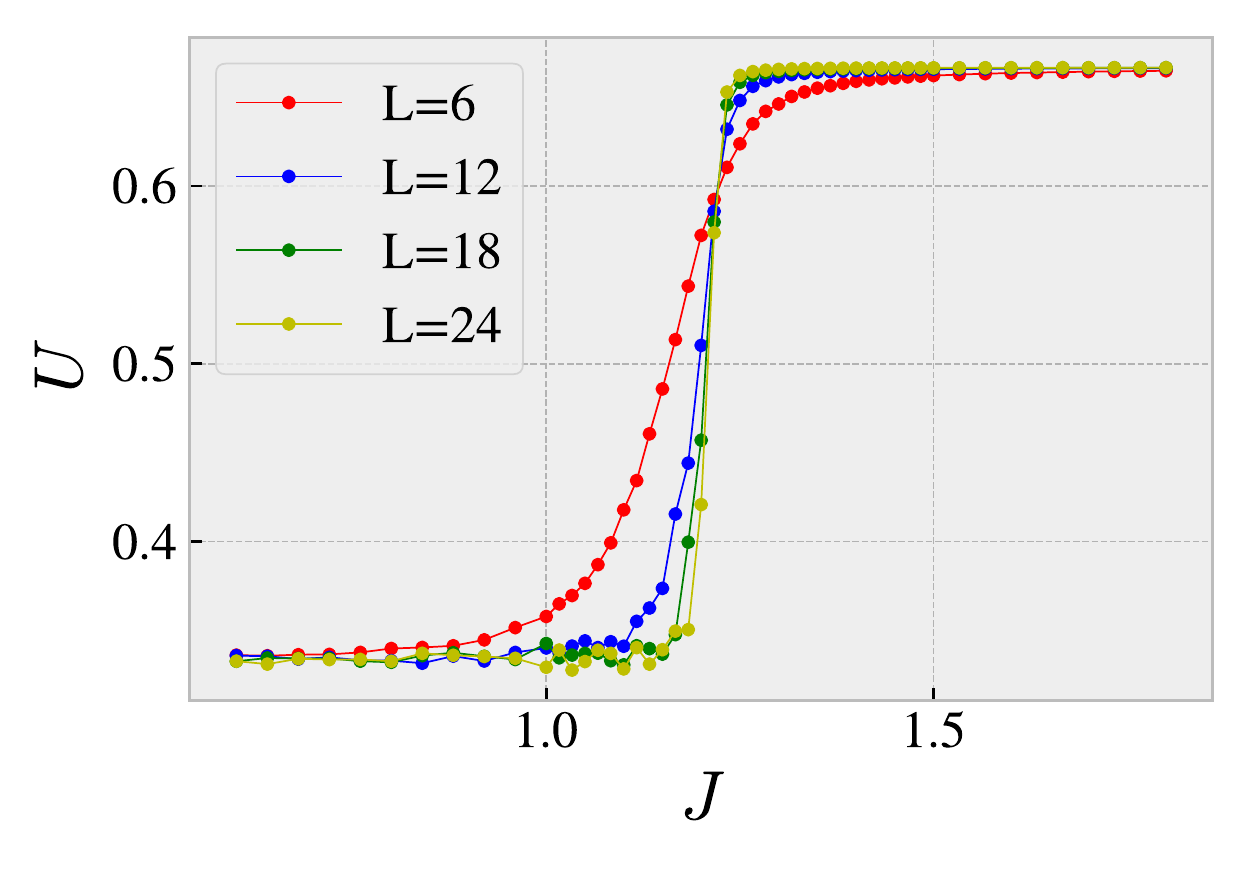}}
(\textbf{b})
\raisebox{-1.1in}{
  \includegraphics[width=.2\textwidth]{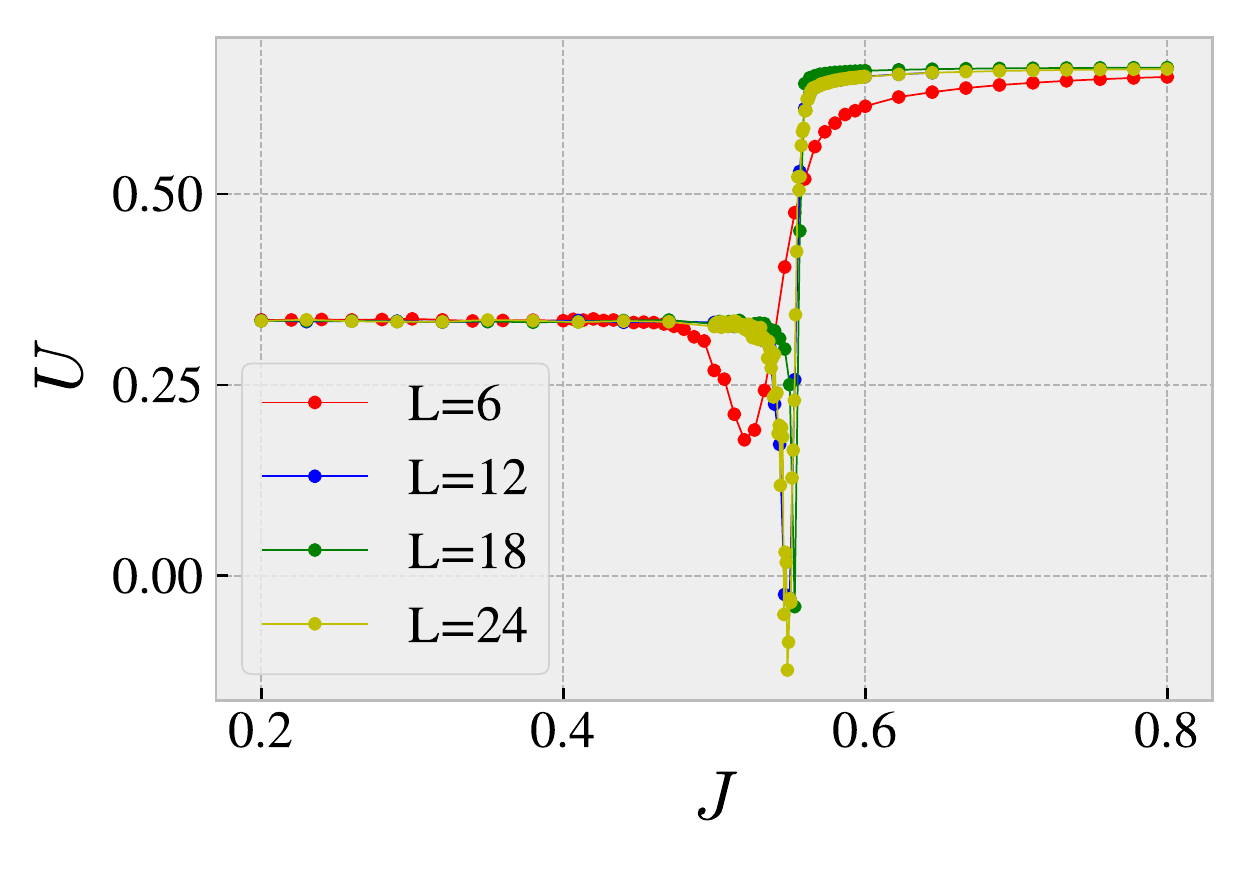}
}
(\textbf{c})
\raisebox{-1.1in}{
  \includegraphics[width=.2\textwidth]{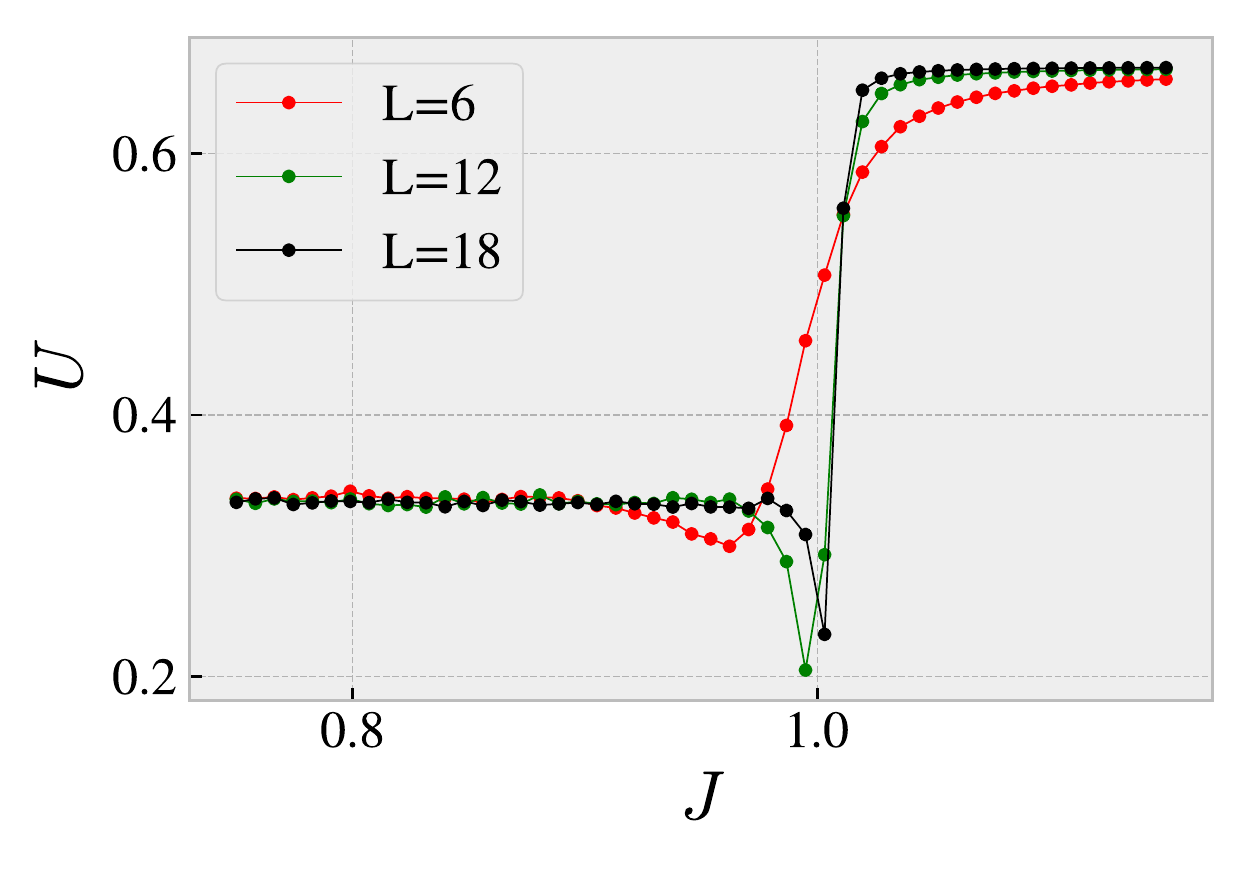}
  }
(\textbf{d})
  \raisebox{-1.1in}{
  	\includegraphics[width=.2\textwidth]{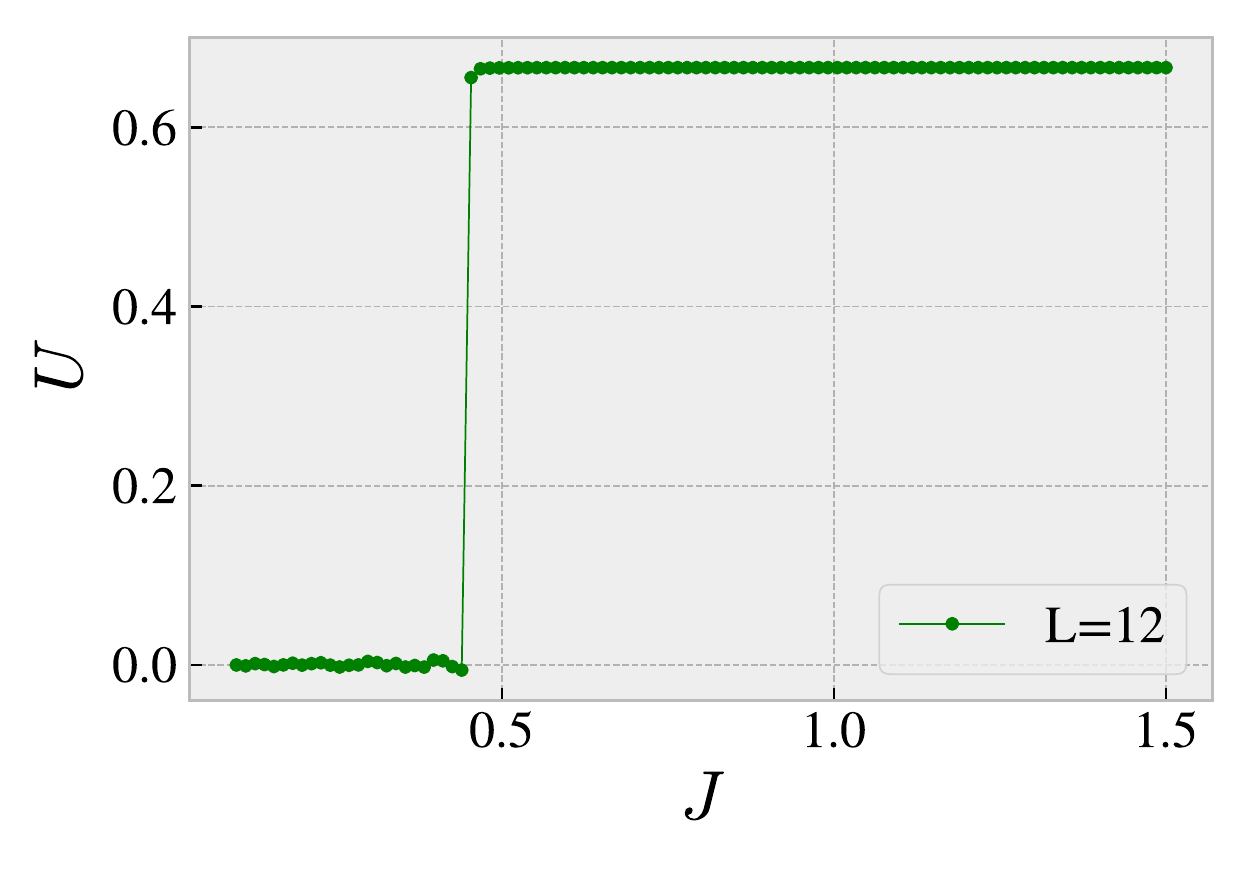}
  }
	\caption{Examples of the Binder ratio as function of system size: (\textbf{a}):  A typical example of a second order transition from confining to Higgs for $O(2)/Z_2$ at $K=0$, (\textbf{b}): An example of a first order transition for $O(2)/Z_2$ at value of $K$ in the range of first order - "Huse-Leibler" line emanating from the tricritical point (\textbf{c}):  An example of the Binder ratio for a weak first order transition, showing its behavior for $O(2) \cross O(2) / Z_2$ for  $K=0$ as function of $J$. (\textbf{d}) An example of the Binder ratio for $Z_2 \cross Z_2 \cross Z_2 / Z_2$ along $K=0$ with a clear indication of a second order phase transition.}
			\label{weakfirstorder}
\end{figure*}

We used several quantities in order to determine the order of a phase transition. For the $O(2)$ matter fields we tracked  the  local nematic magnetization $m=\langle |e^{2i\theta_i}|\rangle$ and local registry order parameter $R=\langle \theta^1_i \cdot \theta^2_i \rangle$. For $Z_2$ matter fields we only encountered the local registry order parameter. Note that formally the $O(2)\cross O(2)/Z_2$ registry order parameter is non-local in that it is the nearest neighbor difference. In practice this implies also a local order parameter, which is easy to understand after performing a ``block-spin'' averaging RG-step. We also measured  the specific heat computed as $C_V = \frac{1}{d^3}(\langle E^2 \rangle -\langle E \rangle)$. In order to see the exact point of phase transition we employed the Binder ratio defined as: 
\begin{equation}
	U = \frac{1}{2} \left[3 - \frac{\langle m^4\rangle}{\langle m^2 \rangle^2}\right] \mathperiod
\end{equation}
In the $\lim_{T \to 0} U = 1$ while for $\lim_{T \to \infty} U = 0$. Since this ratio is dimensionless, plotting curves of different sizes clusters their intersection point which represents the exact value of the phase transition. In our case shape of the Binder curve is more important than the exact intersection point. A smooth transition from 0 to 1 in a sigmoid fashion indicates a second order transition while a sharp dip that diverges with system size indicates a first order behavior. Especially the transition associated with the small $K$ regime of the $O(2) \cross O(2) / Z_2$ is quite weakly first order, and this is manifested by a Binder ratio dip that does not diverge with the system size, see e.g. and Fig.\ref{weakfirstorder}(\textbf(c)). When going to larger system sizes, dip in the Binder curve become very narrow. In that case, refining values of $J$ in that regions helps capture it, otherwise the peak is easily missed.  

\begin{figure*}[!h]
	\centering
	\includegraphics[scale=.7]{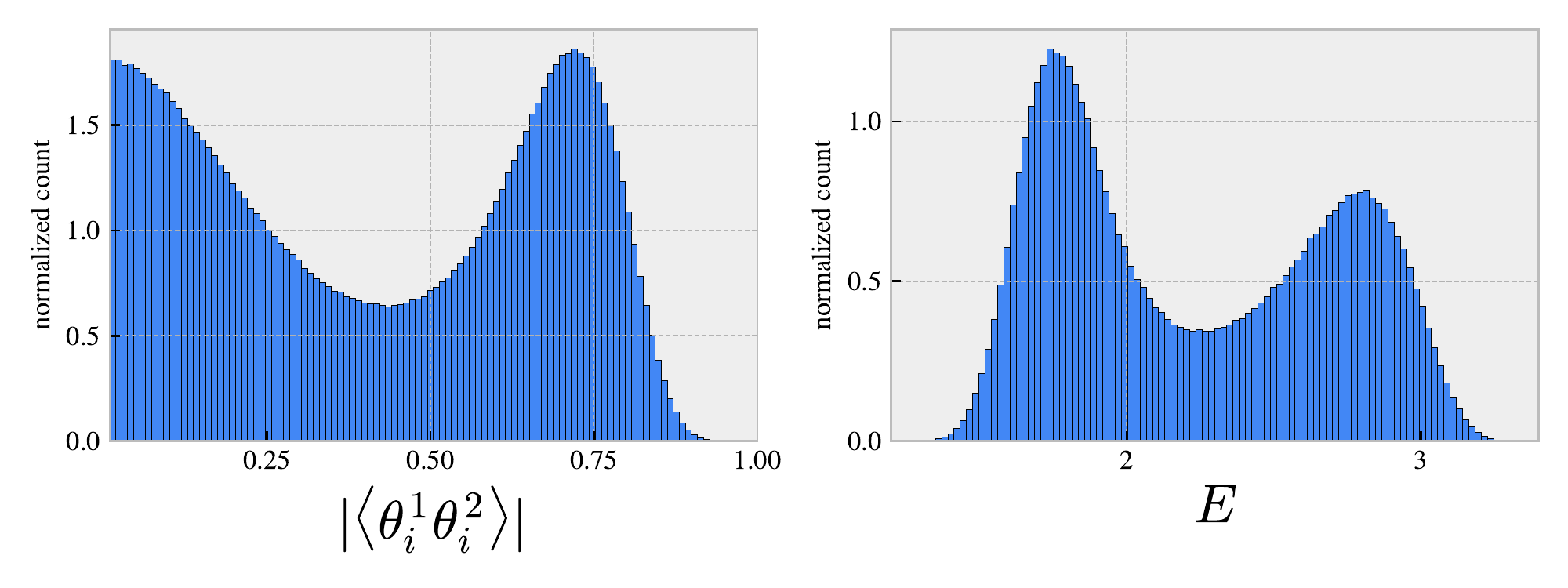}
	\caption{Example of average order parameter ($\langle \theta^1_i \cdot \theta^2_i \rangle$) and energy ($E$) distributions values during a single run of a simulation, on the ``Huse-Leibler'' first order line $J=0.29, K=0.55$ , for $Z_2 \cross Z_2 / Z_2$ model with the usual $J=J_1=J_2$ and $J_L^{12}=0$. We see the appearance of two peaks in these distributions indicating the coexistence of two phases.}
	\label{distributions}	
\end{figure*}

Besides the Binder ratio we also inspected histograms of values occurred during the run of a simulation for an average spatial order parameter and the local energy in order to observe the detailed behavior around the critical point where more than one peak signals the phase separation associated with first order transitions. Histogram are created by counting the occurrences of observed value in specific predetermined bins. Even with fluctuations happening during the run of a simulation, these graphs reveal where are the points of most concentrations around which quantity varies. In most simulations size of the single bin was $10^{-5}$ in single unit of observed quantity, which helps in a resolution of very closely placed peaks. As an example of usefulness of distribution histograms in determining the order of a phase transition we present ``registry'' and energy distributions for $Z_2 \cross Z_2 / Z_2$ model for single value of $K=0.55$ in the regime of ``Huse-Leibler'' first order line emanating from the tricritical point, Fig.\ref{distributions}.

\begin{figure*}[!h]
	\centering
	\includegraphics[scale=.7]{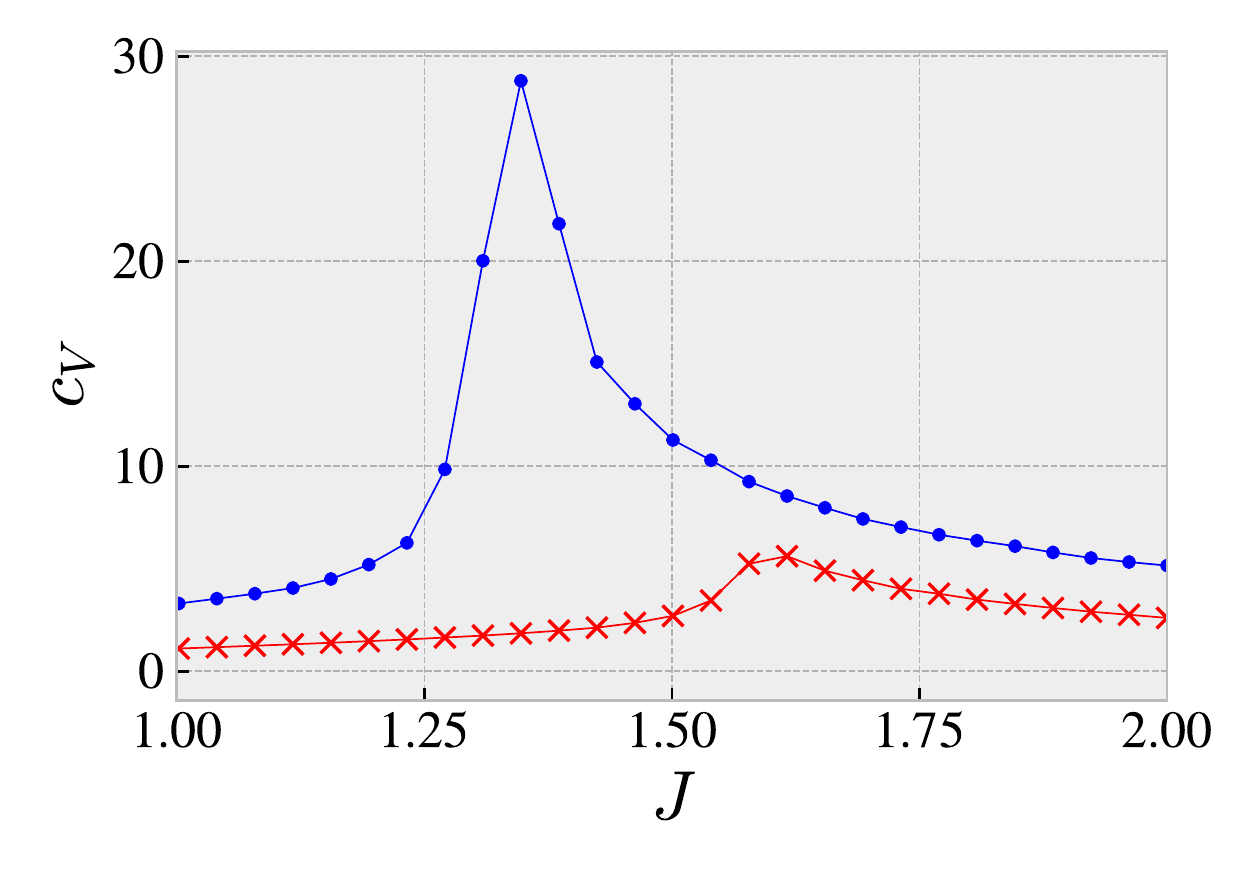}
	\caption{Example of the specific heat $c_V$ for $O(2) \cross O(2) / Z_2$ (blue-dots) and $O(2)/Z_2$ (red-cross) along the $K=0$ slice (Higgs to confining transition). These two models have a different order of a phase transition which might be seen from the amplitude of a $c_V$ divergence. Usually sharper and bigger divergences indicate the first order has occurred. However, one has to be careful due to finite size effects, and for this reason we also study the Binder cumulants. Better use for this graphs is in roughly locating where the transition happens. This information can be used to run the simulation on a more finely spaced grid around the transition in order to get better convergence and more precise point of transition using finite size methods.}
	\label{cvExample}	
\end{figure*}

All the phase diagrams presented in this paper are graphs of specific heat. But for determining the precise nature of the phase order that is not enough. As it can be seen from raw data graphs absolute values of specific heat can be an indicator of the order, but can't be completely trusted, because these values depend on the model and are not universal. We only used specific heat as an indicator of where the transition might be Fig.\ref{cvExample}, but analysis using Binder and histograms were done to determine the order of the transition.


\newpage

\vspace{0mm}
\begin{thebibliography}{99}
\vspace{0mm}

\bibitem{Wegner} F.J. Wegner, J. Math. Phys. {\bf 12}, 2259 (1971).
\bibitem{Kogut} J.B. Kogut, Rev. Mod. Phys. {\bf 51}, 659 (1979).
\bibitem{Wen} X.G. Wen, Phys. Rev. B {\bf 44}, 2664 (1991).
\bibitem{Bais} F.A. Bais, P. van Driel and M. de Wold Propitius , Phys. Lett. B {\bf 280}, 63 (1992). 
\bibitem{Sondhi} T.H. Hansson, V. Oganesyan and S.L. Sondhi, Ann. Phys. {\bf 313}, 497 (2004).
\bibitem{visons} T. Senthil and M.P.A. Fisher, Phys. Rev. B {\bf 62}, 7850 (2000).
\bibitem{stripefrac} J. Zaanen, O.Y. Osman, H.V. Kruis, Z. Nussinov and J. Tworzydlo, Phil. Mag. B {\bf 81}, 1485 (2002); 
J. Zaanen and Z. Nussinov, J. Phys. IV France {\bf 12}, Pr9-245-250 (2002) (arXiv:cond-mat/0209437).
\bibitem{stripefracdemler} Y. Zhang, E. Demler and S. Sachdev, Phys. Rev. B{\bf 66}, 094501 (2002). 
\bibitem{Kitaev} A.Y. Kitaev, Ann. Phys. {\bf 303}, 2 (2003).  
\bibitem{toner95} P.E. Lammert, D.S. Rokhsar and J. Toner, Phys. Rev. Lett. {\bf 70}, 1650 (1993).
\bibitem{nonabnematic} K. Liu, J. Nissinen, R.J. Slager, K. Wu and J. Zaanen, Phys. Rev. X {\ bf 6}, 041025 (2016). 
\bibitem{Fradkinshenker} E. Fradkin and S.H. Shenker, Phys. Rev. D {\bf 19}, 3682 (1979).
\bibitem{HuseLeibler} D.A. Huse and S. Leibler. Phys. Rev. Lett. {\bf 66}, 437 (1991).  
\bibitem{Babaev1} E. Babaev, A Sudbo and N.W. Ashcroft, Nature {\bf 431}, 396 (2004).
\bibitem{Babaev2} E. Babaev and N.W. Ashcroft, Nature Physics {\bf 3}, 530 (2007). 
\bibitem{Babaev3} E. Babaev, J. Carlstrom and M. Speight, Phys. Rev. Lett. {\bf 105}, 067003 (2010). 
\bibitem{Senthil} T. Senthil and O. Motrunich, Phys. Rev. B {\bf 66}, 205104 (2002).
\bibitem{nonabnematic1} J. Nissinen,  K. Liu, R.J. Slager, K. Wu and J. Zaanen, Phys. Rev. E {\bf 94}, 022701 (2016).
 \bibitem{nonabnematic2} K. Liu, J. Nissinen, R.J. Slager, J. de Boer and J. Zaanen, Phys. Rev.E {\bf 95}, 022704 (2017).
 \bibitem{nematic2D}  K. Liu, J. Nissinen, Z. Nussinov,  R.J. Slager, K. Wu and J. Zaanen, Phys. Rev. B {\bf 91}, 075103 (2015).
 \bibitem{PolyakovQED} A.M. Polyakov, Nucl. Phys. B {\bf 120}, 429 (1977);  Phys. Lett. B {\bf 59}, 85 (1975).
 \bibitem{sandvik} S. Jin, A. Sen and A.E. Sandvik, Phys. Rev. Lett. {\bf108}, 045702 (2012).
\end{thebibliography}
\end{document}